\documentclass[reprint,groupedaddress,preprintnumbers,amsmath,amssymb,aps,prab]{revtex4-2} 

\usepackage{graphicx}
\usepackage{dcolumn}
\usepackage{bm}
\RequirePackage[colorlinks,linkcolor=blue,urlcolor=blue,bookmarksnumbered=true]{hyperref}
\RequirePackage[figure]{hypcap}
\RequirePackage[capitalise,nameinlink]{cleveref}
\usepackage{GRAF}
\usepackage[Large]{subfigure}

\usepackage{ifthen}
\newcommand{\Ea}{E_\mathrm{acc}}
\newcommand{\Rres}{R_\mathrm{res}}
\newcommand{\Rbcs}{R_\mathrm{BCS}}
\newcommand{\Ploss}{P_\mathrm{loss}}
\newcommand{\Bext}{B_\mathrm{ext}}

\begin{document}

\title{Improved high-gradient performance for medium-velocity superconducting half-wave resonators: Surface preparation and trapped flux mitigation}%

\author{Yuting Wu}
\author{Kenji Saito}
\email{Corresponding author: saito@frib.msu.edu}
\author{Alex Taylor}
\author{Andrei Ganshyn}
\author{Chris Compton}
\author{Ethan Metzgar}
\author{Kyle Elliott}
\author{Laura Popielarski}
\author{Sam Miller}
\author{Sang-hoon Kim}
\author{Spencer Combs}
\author{Taro Konomi}
\author{Ting Xu}
\author{Walter Hartung}
\author{Wei Chang}
\author{Yoo-Lim Cheon}
\affiliation{Facility for Rare Isotope Beams, Michigan State University, East Lansing, MI, USA}

\date{\today}

\begin{abstract}
A development effort to improve the performance of superconducting radio-frequency half-wave resonators (SRF HWRs) is underway at the Facility for Rare Isotope Beams (FRIB), where 220 such resonators are in operation. Our goal was to achieve an intrinsic quality factor ($Q_0$) of $\geq 2\times 10^{10}$ at an accelerating gradient ($\Ea$) of 12~MV/m. FRIB production resonators were prepared with buffered chemical polishing (BCP).
First trials on electropolishing (EP) and post-EP low temperature baking (LTB) of FRIB HWRs allowed us to reach higher gradient (15 MV/m, limited by quench) with a higher quality factor at high gradient, but $Q_0$ was still below our goal.  Trapped magnetic flux during the Dewar test was found to be a source of  $Q_0$ reduction. Three strategies were used to reduce the trapped flux: (i) adding a local magnetic shield (LMGS) to supplement the ``global'' magnetic shield around the Dewar for reduction of the ambient magnetic field; (ii) performing a ``uniform cool-down'' (UC) to reduce the thermoelectric currents; and (iii) using a compensation coil to further reduce the ambient field with active field cancellation (AFC). The LMGS improved the $Q_0$, but not enough to reach our goal. With UC and AFC, we exceeded our goal, reaching $Q_0 = 2.8\times10^{10}$ at $\Ea$ = 12 MV/m.

\end{abstract}

\maketitle


\newlength{\figLHwidth}%
\newlength{\figXLwidth}%
\newlength{\figHwidth}%
\newlength{\figHLwidth}%
\newlength{\figHLLwidth}%
\newlength{\figFwidth}%
\newlength{\figSwidth}%
\newlength{\figKGwidth}%
\ifthenelse{\lengthtest{\columnwidth=\textwidth}}%
{
\setlength{\figLHwidth}{0.675\columnwidth}%
\setlength{\figXLwidth}{0.875\columnwidth}%
\setlength{\figHwidth}{0.5\textwidth}%
\setlength{\figHLwidth}{0.48\textwidth}%
\setlength{\figHLLwidth}{0.45\textwidth}%
\setlength{\figFwidth}{0.41\textwidth}%
\setlength{\figSwidth}{0.85\columnwidth}%
\setlength{\figKGwidth}{0.8\columnwidth}%
}%
{
\setlength{\figLHwidth}{\columnwidth}%
\setlength{\figXLwidth}{\columnwidth}%
\setlength{\figHwidth}{\columnwidth}%
\setlength{\figHLwidth}{\columnwidth}%
\setlength{\figHLLwidth}{\columnwidth}%
\setlength{\figFwidth}{\columnwidth}%
\setlength{\figSwidth}{\columnwidth}%
\setlength{\figKGwidth}{\columnwidth}%
}

\section{Introduction}

Superconducting radio-frequency (SRF) cavities have grown to play an increasingly important role
in particle accelerator technology.  The 60-year anniversary of the advent of research and development in SRF cavities was commemorated at the last international conference on RF superconductivity in 2023 \cite{SRF2023:proc}.  Progress in the field in the past decades has well-been documented in the literature, as seen, for example, in Refs.~\cite{CRYO18:465to471, IEEETNS30:3309to3312, ARNPS43:435to686, padamsee1998rfsca, PPNP49:155to244, hsp2009:rfsc:sta, IPAC2012:MOYBP01, SCST:30:034004, hsp2023:srfta:saet}.  Initially, SRF cavities were used primarily to accelerate electrons and positrons traveling near the speed of light ($\beta = v/c \approx 1$) and ions traveling at low speeds ($\beta \ll 1$); recent years have witnessed an expansion in the application of SRF to intermediate velocities for proton and ion acceleration.

Improved performance of SRF accelerators has been a long-standing goal of the SRF community, with considerable effort aimed toward operating at higher accelerating gradient ($\Ea$) and with higher intrinsic quality factor ($Q_0$); both of these having a crucial impact on the performance, capital cost, operating cost, and real estate requirements of SRF accelerators.

\subsection{Low-field performance}

At low field, $Q_0$ is inversely proportional to the RF surface resistance $R_s$, which is the sum of a temperature-independent residual surface resistance $\Rres$ and a temperature-dependent term $\Rbcs$ named for the Bardeen-Cooper-Schrieffer (BCS) model of superconductivity.  The BCS contribution increases quadratically with the RF frequency.  As a result, lower-frequency cavities (of order 80 MHz) typically operate at higher temperature (4.2~K or above) and higher-frequency cavities (of order 1.3~GHz and up) operate at lower temperature (typically 2~K).  At intermediate frequencies, cavities may be operated at 4.2~K with better cryogenic plant efficiency or at 2~K with lower power dissipation and lower $\Rbcs$ than for the high-frequency case.  In the latter scenario, the contribution from $\Rres$ relative to $\Rbcs$ is more significant than for the high-frequency case.

\subsection{Magnetostatic flux trapping and thermo-electric currents: Basic observations\label{S:TrapBasic}}

Although an ideal superconductor expels magnetic fields, real-world SRF cavities tend to trap the magnetic flux component normal to the surface \cite{EPAC1992:1295}.
The trapped flux decreases the low-field $Q_0$.  Hence minimization of the residual magnetostatic field is
important for high-performance cavities.  As complete elimination of the residual field may not be practical, minimization of the contribution to the residual resistance from trapped flux (``sensitivity'') is desirable as well.

The Seebeck effect adds an additional complication---current loops may be set up if dissimilar metals and temperature gradients are present when the cavity transitions from normal conducting to superconducting; the magnetic field associated with the current loop may then be trapped, resulting in a decrease in $Q_0$ even without a residual ambient field.  Reductions in cavity performance have been observed (i) in niobium cavities with an inner layer of Nb$_3$Sn due to thermo-currents resulting from temperature gradients during fast cool-down \cite{SRF1987:E04}; (ii) due to thermo-currents induced in a cavity quench at high RF field \cite{SRF1997:B06, SRF1999:TUP005}; and (iii) due to thermo-currents induced during fast cool-down of niobium cavities with titanium jackets \cite{SRF2009:TUPPO053, PRSTAB16:102002}.  We will return to this discussion in \cref{S:TrapMitig}. 

\subsection{High-field performance and \texorpdfstring{$Q$}{Q}-slope\label{S:HiFldIntro}}

Mechanisms by which the quality factor may decrease as the accelerating field increases become more
problematic as we try to operate cavities at higher gradients.  A``medium-field $Q$-slope'' (MFQS) and a ``high-field $Q$-slope'' (HFQS) may be present.  The MFQS is typically more severe at 4.2~K and less so at 2~K\@.

One model to explain MFQS is the thermal feedback (TF) model, which considers
the increase in the temperature of the inner surface of a cavity relative to the temperature of the liquid helium bath as a function of the RF power dissipation. 
This model is described in the literature, for example, in Section~3.3.2 of Ref.~\cite{hsp2009:rfsc:sta}.  
Other models have been proposed as well, for example a hysteresis loss model~\cite{ciovati2006analysis}, a two-fluid, multi-mechanism model~\cite{PRSTAB14:101002}, and a density-of-states-smearing model \cite{APL104:092601}; the TF model and other mechanisms to produce an RF-field-dependent surface resistance are an area of active study,
as exemplified by Refs.~\cite{ciovati2006analysis, PRSTAB14:101002, APL104:092601, SST27:085004, PRAB27:092001}.

In addition to intrinsic $Q$-slope, field emission (FE) may further undermine the performance at high fields as a result of electrons being extracted from the cavity walls, accelerated by the RF field, and impacting the cavity surface again.  FE can produce Bremsstrahlung X-rays and, in severe cases, a decrease in $Q_0$.  The risk of field emission can be reduced by minimizing particulate contamination on the cavity surfaces.

\subsection{Surface preparation\label{S:EPintro}}

An important step in the preparation of high-performance SRF cavities is the removal of the ``damage'' layer after cavity fabrication.  For bulk  niobium cavities, this was traditionally done via buffered chemical polishing (BCP\@).  Disadvantages of BCP are that it does not reduce the surface roughness, and BCP'ed cavities tend to show a steep HFQS\@.  Electro-polishing (EP) is an alternative to BCP\@.
Early EP development for Nb cavities was done in the 1970s at Siemens \cite{PLA37:139to140, IEEETNS20:68to78}.
EP was then was applied to low-$\beta$ cavity sub-assemblies at Argonne \cite{IEEETNS24:1147to1149}
and full high-$\beta$ cavities at KEK \cite{SRF1989:D02, SRF1989:G18}.
EP can produce a smoother surface than BCP \cite{PAC2003:ROAA002}, but adds complexity.
The application of EP for SRF cavities has expanded significantly in recent years, as documented in the literature, for example
Section 7.4 of Ref.~\cite{hsp2009:rfsc:sta},
Section 8.1 of Ref.~\cite{hsp2023:srfta:saet}, and Refs.~\cite{PAC2003:ROAA002, SST30:043001}.
A notable advance in recent years was the application of EP to fully-formed low- and medium-$\beta$ cavities at Argonne \cite{SRF2011:WEIOA03, IPAC15:WEPTY010}.

\subsection{Basic thermal treatments\label{S:LTBintro}}

For high-performance Nb cavities, heat treatment at 600 to 800~$^\circ$C is typically done after bulk BCP or bulk EP
to remove hydrogen from the cavity walls and mitigate the formation of lossy hydrides at low temperatures, as described, for example, in Section 7.5 of Ref.~\cite{hsp2009:rfsc:sta}.

After final light BCP or EP, in-situ low-temperature baking (LTB, typically at 120$^\circ$C for 48 hours) may be done.
The effects of LTB have been observed and studied by a number of research teams for both higher-frequency cavities~\cite{saito1999high, SRF1999:TUP015, SRF1999:TUP044, ciovati2004effect, PRSTAB13:022002, PRSTAB16:012001, ieeetas23:1to6} and lower-frequency cavities \cite{SRF2013:THP016, NIMA868:82, SRF2023:TPUTB035, SRF2015:WEBA04, SRF2017:THXA08, SRF2015:WEA1A03, FEMAT3:1244126}.  We provide a brief overview in this section and refer the reader to the literature for more in-depth surveys of LTB findings and their impact on cavity performance, for example Chapter 5 of Ref.~\cite{hsp2009:rfsc:sta} and Chapter 6 of Ref.~\cite{hsp2023:srfta:saet}.

LTB can be understood as a method to modify the surface properties of an SRF cavity without changing the bulk properties. 
At low field, LTB has been shown to reduce $\Rbcs$ via reduction of the mean free path near the surface (within the London penetration depth).
This can explain the usefulness of LTB for high-frequency cavities operating at 2~K (where $\Rbcs$ is large due to the high frequency) and for low-frequency cavities operating at 4.2~K (where $\Rbcs$ is large due to the high temperature).  On the other hand, there is less benefit from LTB for intermediate frequencies (of order 320~MHz) with an operating temperature of 2~K, as the temperature is low and the frequency allows for the BCS term to be significantly lower than the high-frequency case.

LTB has been observed to affect the $Q$-slope, in addition to the low-field $Q_0$.  Use of LTB after EP is known to provide effective mitigation of HFQS for high-frequency elliptical cavities thanks to studies at KEK and other laboratories in the 1990s~\cite{saito1999high, SRF1999:TUP015, SRF1999:TUP044}.
EP and LTB are now standard steps for preparation of high-performance $\beta = 1$ cavities operating at 1.3~GHz~\cite{TTC2008:05, SRF2011:TUPO015}.  Follow-up studies and observations have provided additional insight into the effects of LTB, as seen, for example, in Refs.~\cite{ciovati2004effect, PRSTAB13:022002, PRSTAB16:012001, ieeetas23:1to6}.  This includes modeling of the $Q$-slope \cite{ciovati2004effect}; studies of the impact of the grain size of the parent Nb sheet \cite{PRSTAB13:022002}; studies of performance as a function of depth of penetration of surface impurities \cite{PRSTAB16:012001}; and studies of the relationship between LTB and flux trapping \cite{ieeetas23:1to6}.

For low-frequency cavities operating at $\sim 4.2$~K, the benefits of LTB are generally more due to an improvement in MFQS, more so than to the increase in the low-field $Q_0$; we will return to this point in \cref{S:LowBeta}.

There has been some additional experience with intermediate-$\beta$ multi-cell elliptical cavities at intermediate frequencies (of order 650 MHz) and the impact of LTB on their performance \cite{PRAB24:112003, NIMA1059:168985, martinello2021q}.

\subsection{Doping and other alternative thermal treatments\label{S:Dopeintro}}

Additional measures for surface modification have been explored in recent years, including ``medium-temperature baking'' (MTB)  \cite{PRAPPL13:015024, SST34:095005, SRF2021:FROFDV01, PhysicaC599:1354092, SRF2023:MOPMB030} and nitrogen doping or infusion \cite{SST26:102001, SST30:094004, PRAB21:032001, PhysOpen5:100034, PhysOpen5:100034}.  These treatments have been observed to produce an ``anti-$Q$-slope'' for 1.3~GHz cavities at intermediate fields. N-doping and MTB have been shown to be beneficial for ILC-type cavities~\cite{NIMA883:143to150}.  N-doping was included as a preparation step for LCLS-II cryomodules; the transition from development to production provided some additional lessons and insights~\cite{NIMA883:143to150}.

N-doping and MTB are being evaluated for lower frequencies, along with LTB \cite{PRAB24:112003, NIMA1059:168985, martinello2021q}.
Initial studies for elliptical cavities with $\beta \sim 0.65$ and frequency $\sim 650$~MHz found that N-doping can increase the quality factor, but an anti-$Q$-slope has not been observed \cite{NIMA1059:168985, martinello2021q}.  One study on a HWR at 162.5~MHz found that nitrogen infusion mitigated the $Q$-slope at 2~K, again, without anti-$Q$-slope \cite{NIMA937:21to25}.  These results suggest that there may be a frequency dependence in the effects of N-doping or N-infusion, which would be consistent with frequency dependence studies for higher frequencies \cite{PRL121:224801}. 

\subsection{Magnetostatic flux trapping and thermo-electric currents: Studies, mitigations, interdependencies\label{S:TrapMitig}}

Per \cref{S:TrapBasic}, flux trapping, sensitivity to trapped flux, and thermo-currents
may significantly impact the quality factor of an SRF cavity.  These topics have been the object of
a number of studies conducted in parallel with the studies on heat treatment and doping
described above; some examples are included in the reference list.
The studies have been oriented toward both high-$\beta$ elliptical
cavities \cite{IEEETMAG:1620to1623, SRF1995:D02, SRF1999:TUP005,
JAP115:184903, APL105:234103, JAP118:044505, JAP119:213903,
PTEP2016:053G01, APL109:062601, APL112:072601, PRAB23:123101} and
lower-$\beta$ coaxial cavities \cite{SRF2017:WEXA06, PRAB24:083101,
SRF2023:MOPMB023}.  
As they are typically operated at higher temperature, trapped flux
is typically a lesser concern for lower-frequency cavities; consequently, flux trapping in
higher-frequency cavities has received more attention.

Aspects studied have included the effect of
cool-down speed and temperature gradients \cite{JAP115:184903,
APL105:234103, JAP119:213903, PTEP2016:053G01, SRF2017:WEXA06,
SRF2023:MOPMB023}; the dependence on frequency \cite{APL112:072601,
SRF2023:MOPMB023}; the impact of cavity orientation
\cite{JAP118:044505, PRAB23:123101, SRF2023:MOPMB023}; the impact of
material properties \cite{PRAB24:083101}, thermal treatments
\cite{ieeetas23:1to6, JAP119:213903, APL109:062601, APL112:072601}, and doping
\cite{APL109:062601, APL112:072601}; the differences between bulk Nb
cavities and Nb-film-on-Cu cavities \cite{IEEETMAG:1620to1623,
SRF1995:D02, SRF1997:B05, SRF2017:WEXA06}; the impact of cavity
fabrication methods \cite{SRF2017:WEXA06}; and the dependence on the
RF field amplitude \cite{APL112:072601}.  More in-depth information can be
found in the literature, including Chapter 5 of
Ref.~\cite{hsp2023:srfta:saet}.  High quality factors have been shown to be
possible with careful cavity preparation and careful mitigation of flux
trapping \cite{JAP115:184903, APL105:234103, JAP119:213903}, though some of the mitigation methods are
not easily applied to improved-performance SRF cryomodules.
Slow cool-down through the transition temperature has been found to be
beneficial to mitigate thermo-currents in the presence of dissimilar materials \cite{SRF1987:E04, SRF2009:TUPPO053, PRSTAB16:102002, SRF2017:WEXA06}.
However, fast cool-down has been found to be beneficial by helping to provide flux expulsion,
especially after MTB or N-doping and LTB \cite{JAP115:184903, APL105:234103}.  As a result of these considerations,
a sophisticated cool-down method was needed for LCLS-II cryomodules with
N-doped cavities~\cite{IOPCSMSE1327:012004}.

\subsection{The FRIB Linear Accelerator}

The Facility for Rare Isotope Beams (FRIB) at Michigan State University (MSU) operates a superconducting linear accelerator for user experiments with heavy ions~\cite{wei2012frib, mpla37:2230006}. The linac uses quarter-wave resonators (QWRs, 80.5 MHz) and half-wave resonators (HWRs, 322 MHz) for ion acceleration, with a total of 324 resonators in 46 cryomodules~\cite{xu2021completion}.  \Cref{fig:HWR:DWGnMASS} shows a drawing of $\beta = 0.53$ HWR and a photograph of HWRs being assembled into a cold mass.  Cavity design parameters can be found in a previous paper \cite{NIMA1014:165675}.  
FRIB production cavities were treated with BCP at MSU's etching facility \cite{LINAC2014:THPP046} and hydrogen-degassed in a furnace at 600~$^\circ$C for 10 hours \cite{LINAC2012:MOPB071}; LTB was not included as a routine preparation step.
Cavity certification testing \cite{NIMA1014:165675} and cryomodule certification testing \cite{NAPAC2019:WEPLM73} were done at MSU\@.

Long-term linac operation with high availability and high reliability is required to fulfill the facility's science mission. As the linac performance may degrade over time, we anticipate the need for high-performance spare and refurbished cryomodules over the facility lifetime.

\begin{figure}
   \centering
    \GRAFwidth[0.4\textwidth]{296}{434}
    \GRAFlabelcoord{261}{20}
    \GRAFlabelbox{35}{25}
    \incGRAFboxlabel{HWR}{(a)}
\ifthenelse{\lengthtest{\columnwidth=\textwidth}}
{\relax}%
{\\[2ex]}
    \GRAFwidth[\figHwidth]{504}{599}
    \GRAFlabelcoord{444}{0}
    \GRAFlabelbox{60}{45}
    \incGRAFboxlabel{s53_rails_clean}{(b)}
       \caption{(a) Isometric sectional view of a FRIB $\beta$ = 0.53 HWR.  The helium jacket is shown in green.  (b) Photograph of a $\beta = 0.53$ cold mass during clean-room assembly.}
   \label{fig:HWR:DWGnMASS}
\end{figure}

\subsection{HWR performance improvement efforts for FRIB\label{S:TransProc}}

The above considerations led us to begin a development effort in 2022 oriented toward improving the high-field performance of FRIB HWRs. Our tangible goal was to build a spare medium-velocity ($\beta = 0.53$) HWR cryomodule which can operate at higher accelerating gradient ($\Ea$) with a higher intrinsic quality factor ($Q_0$). Four objectives were planned for this effort: (1) achievement of higher performance using EP and LTB; (2) increasing the $Q_0$ to $2\times10^{10}$ at $\Ea$ = 12~MV/m by reducing the trapped magnetic flux; (3) development of HFQS-free BCP; and (4) development of nitrogen doping during EP (``wet N-doping'').  We successfully completed Objective 1 and Objective 2 in December 2023. The results of Objective 1 were reported in a previous paper~\cite{saito2023development}; we will provide some additional results and some reinterpretation in \cref{S:summ1}. The main focus of this paper will be to report the results of Objective 2.  Work on Objective 3 and Objective 4 is ongoing, with first results for Objective 4 having been reported in a recent conference paper \cite{LINAC2024:TUAA002}. 

\subsection{Other medium-velocity cavity efforts}

A number of other groups have developed or are developing SRF resonators similar to the FRIB $\beta = 0.53$ HWR for other projects.  Recent examples include single-spoke resonators for PIP-II at Fermilab~\cite{parise2023pip} and MYRRHA~\cite{moretti2023fabrication} as well as double-spoke resonators for ESS~\cite{elias2017testing}, the China ADS~\cite{jiang2018development}, and the China SNS upgrade~\cite{zhou2021development}.  Mostly BCP has been used for these spoke resonators, at least in the prototyping phase; our work on improving the performance of FRIB HWRs hence may have relevance for other projects, although the complexity of some spoke resonator designs, especially double-spoke versions, may make the application of EP more challenging.  As foreshadowed in \cref{S:LTBintro},
LTB is not typically included as a preparation step for 2~K operation of such cavities; though LTB has been considered as a potentially useful addition for operation of spoke cavities at higher temperature \cite{SRF2023:THIAA04}.

\subsection{Lower-velocity cavity efforts\label{S:LowBeta}}

Development of HWRs by other groups has been done as well, though typically at lower frequencies and for lower $\beta$ values; examples include HWRs developed at Argonne \cite{SRF2015:WEBA05}, at IMP \cite{SRF2013:THP016}, for RISP \cite{NIMA868:82}, and for IFMIF \cite{SRF2023:TPUTB035}, as well as operational HWRs for SARAF \cite{RUPAC2012:WEBCH01}.  Our findings may have some relevance to such cavities, in spite of the differences in size, shape, and operating frequency.  Some LTB results for lower-frequency HWRs have been reported previously, albeit for cavities prepared with BCP \cite{SRF2013:THP016, NIMA868:82, SRF2023:TPUTB035}; HWRs prepared with EP at Argonne did not receive LTB \cite{SRF2015:WEBA05}.

For even lower velocities, QWRs are typically used at even lower frequency, typically operating at $\sim 4.2$ K, and typically prepared with BCP\@.  
LTB is often used for QWRs, and is considered helpful to soften multipacting barriers.  An LTB step was included for the higher-$\beta$ Spiral2 QWRs~\cite{SRF2015:WEBA04, SRF2017:THXA08} and was found to be beneficial for ISAC-II QWRs~\cite{SRF2015:WEA1A03, FEMAT3:1244126}.  Per \cref{S:LTBintro}, a lessening of the MFQS at $\sim 4.2$~K was seen with LTB in both cases.

There are 104 QWRs in the FRIB linac, operating at 80.5~MHz.  Presently, our EP facility is not able to accommodate QWRs.  However, LTB has been added as a preparation step for production of spare and refurbished FRIB $\beta = 0.085$ QWR cryomodules, as this allows for a higher quality factor at the operating gradient and operating temperature of 4.5~K \cite{qwrbake:JBrown:2025}.  Again, the benefit is primarily due to an improvement in the MFQS.

\section{Background}

\subsection{Methods: FRIB cavity production\label{S:CavProd}}

FRIB cavities were fabricated from high-purity sheet Nb (residual resistivity ratio = RRR $\geq$ 300) by deep drawing and electron beam welding.  Production cavities were produced by industrial partners after jacketing.  After delivery to FRIB, production cavities received bulk BCP, followed by hydrogen degassing and light BCP.  As has become standard practice, to reduce the risk of field emission, cavities receive high-pressure rinsing with ultra-pure water and are assembled to the test stand in a clean room environment.
More detailed information can be found in  Ref.~\cite{saito2023development}.

\subsection{Methods: FRIB cavity testing\label{S:CavTest}}

All FRIB production cavities were cold-tested at FRIB for certification prior to being installed in a cryomodule \cite{NIMA1014:165675}.  Dewar testing was done with the cavities in the same orientation as used for the cryomodule; the helium tank was filled with liquid and the Dewar provided the insulating vacuum, in order to approximate the cryomodule environment.

A dedicated cryogenic plant was used for cold testing of cavities and cryomodules.  For production cavity testing, the time to cool a cavity from room temperature to 4.3~K was typically of order 45 to 90 minutes.

Continuous wave (CW) measurements at 4.3~K are done with continuous filling; conditioning of multipacting barriers is typically done at this time.  We do low-field CW measurements while pumping on the bath to decrease the temperature to the operating goal of 2~K, which allows us to infer the low-field $Q_0$ as a function of temperature. Measurements at 2~K measurements are done with the liquid supply valve closed, which provides a relatively short time window for the measurements.  Measurements are typically started before the bath temperature reaches 2~K, while the bath temperature is still decreasing; at high cavity field, the bath temperature may increase due to the additional power dissipation.  Typically, 2~K measurements are done while increasing the field and are then repeated while decreasing the field.

An ionization chamber is used to monitor X-ray signals during the cold test.
The sensor is located outside the Dewar and inside the radiation shield, about 1 meter from the cavity (Figure 5 in Ref.~\cite{NIMA1014:165675} shows the approximate location).  X-rays are seen in case of field emission; X-rays are also seen when conditioning the highest multipacting barrier.

The FRIB test Dewars have a magnetic shield to reduce the ambient magnetic field, as will be discussed in \cref{S:resid}.  A compensation coil was added to further reduce the ambient field, though this was not used during FRIB cavity production.

\section{Overview: performance improvement efforts\label{S:over}}

FRIB production cavities were all prepared with BCP\@. All of the $\beta= 0.53$ HWRs that were tested to gradients of $> 12$~MV/m showed HFQS, while a majority of them showed some field emission (FE) X-rays at high field~\cite{NIMA1014:165675, saito2023development}.  As described in \cref{S:EPintro} and \cref{S:LTBintro}, EP and LTB are good candidates to mitigate HFQS, and EP could reduce FE as well.

Recently, an EP facility was commissioned at FRIB; this system became operational at the end of 2022~\cite{metzgar2023summary}. The EP facility was designed for both FRIB HWRs and multi-cell elliptical cavities, which are under development for the proposed FRIB energy upgrade \cite{PRAB24:112003, NIMA1059:168985}.  For the present work, we used the FRIB EP facility for surface preparation of FRIB $\beta = 0.53$ HWRs. Objective 1 was successfully completed in May 2023, as will be summarized in \cref{S:summ1}. With EP and LTB, the cavity performance showed a high $Q_0$ without HFQS up to 15 MV/m (limited by quench), as expected.

EP and LTB were not sufficient to reach our $Q_0$ goal, which underscored the need for Objective 2. As discussed in \cref{S:LTBintro}, for intermediate frequencies such as 322~MHz, the contribution from BCS losses is relatively small at 2~K, making the residual losses a larger fraction of the total. Typical values for the FRIB HWR case are $\Rbcs = 0.5$~n$\Omega$ with EP+LTB and $\Rres = 4.5$~n$\Omega$.

Per \cref{S:TrapBasic}, the residual magnetic field is a possible contributor to $\Rres$. The Dewars for FRIB cavity testing are imperfectly shielded, as will be discussed in \cref{S:resid}, with a residual field of 2 to 5~milligauss at the cavity location.  For improved shielding, a local magnetic shield (LGMS) was designed for the $\beta = 0.53$ HWR\@.  Our goal was to reduce the field to $\leq$ 0.1~mG, which should make the power dissipation due to trapped flux negligible. Measurements on the LMGS confirmed that the shielding goal was met at room temperature. The design of the LMGS will be described in \cref{S:lmgs}. The LMGS was installed on a spare FRIB $\beta= 0.53$ HWR (S53-159), which was kept under vacuum by active pumping for 4 months after the Objective 1 tests. An increase in $Q_0$ was seen with the LMGS, but not enough to meet our goal.

Per \cref{S:TrapBasic}, the Seebeck effect is a possible source of losses not mitigated by the LMGS\@. 
In our case, the cavity (niobium), helium jacket (titanium), flanges, and connections for Dewar testing bring together several dissimilar metals, which could produce thermo-electric currents as the cavity is cooled down.  To quantify the contribution from thermo-currents, two cool-down methods were compared: fast cool-down (FC), and uniform cooling (UC), with and without active field cancellation (AFC\@).  UC and AFC together produced the highest $Q_0$ of $2.8\times10^{10}$ at $\Ea$=12 MV/m, as will be described in \cref{S:res}.

\section{Objective 1: EP and LTB---summary, additional results, and reinterpretation\label{S:summ1}}

\begin{figure}[bp]
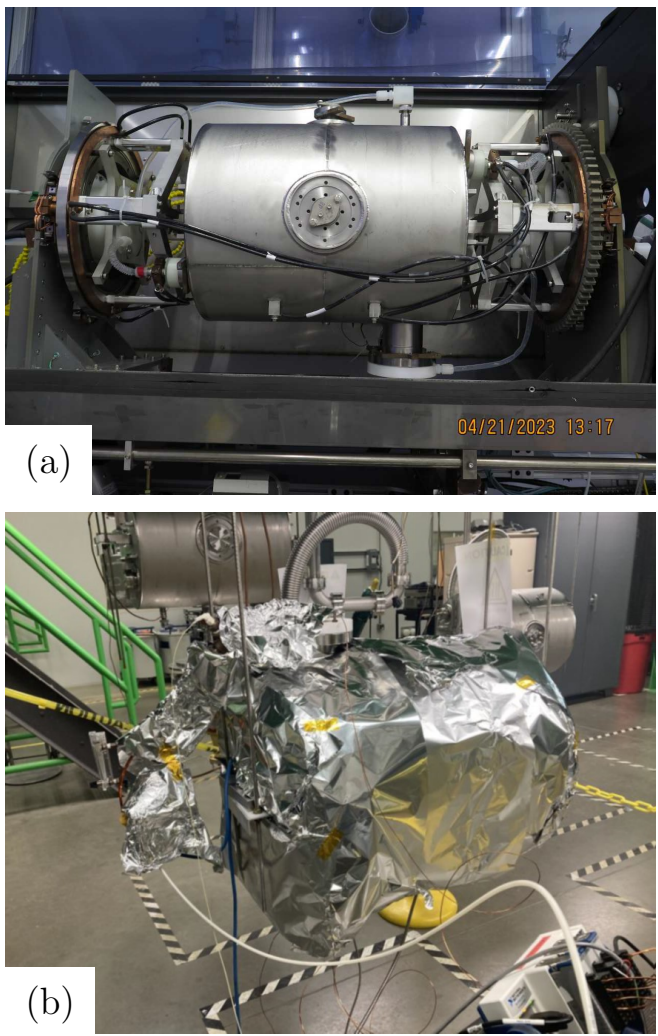

   \centering
    \GRAFwidth[\figHLwidth]{1280}{960}
    \GRAFlabelcoord{0}{0}
    \GRAFlabelbox{140}{120}
    \incGRAFboxlabel{EP}{(a)}
\ifthenelse{\lengthtest{\columnwidth=\textwidth}}
{\relax}%
{\\[1.5ex]}
    \GRAFwidth[\figHLLwidth]{879}{705}
    \GRAFlabelcoord{0}{0}
    \GRAFlabelbox{100}{80}
    \incGRAFboxlabel{s53_159LTB}{(b)}
    \caption{Jacketed $\beta = 0.53$ HWR (a) installed in the horizontal rotational EP system and (b) undergoing a low-temperature bake.}
   \label{fig:HWR:EPnLTB}
\end{figure}

Cavity surface preparation with EP followed by LTB was developed for FRIB $\beta$ = 0.53 HWRs in Objective 1.  Photographs of the EP and LTB stations are shown in \cref{fig:HWR:EPnLTB}.    
Bulk EP (for removal of about 120~$\mu$m total) was done with an acid outlet temperature of $\sim 25^{\circ}$C, as developed previously for ILC cavities.  The final light EP was done with colder acid (outlet temperature $\sim 15^{\circ}$C), unlike the standard ILC cavity preparation.  Whether the colder temperature for final EP impacts the performance of the HWRs is an open question.  LTB was done with the cavity on the test insert, bringing the temperature to $120^\circ$C for 48 hours.
Major results of our study relate to the high-field $Q$-slope~\cite{saito2023development}, and are revisited below.

\begin{figure}[bp]
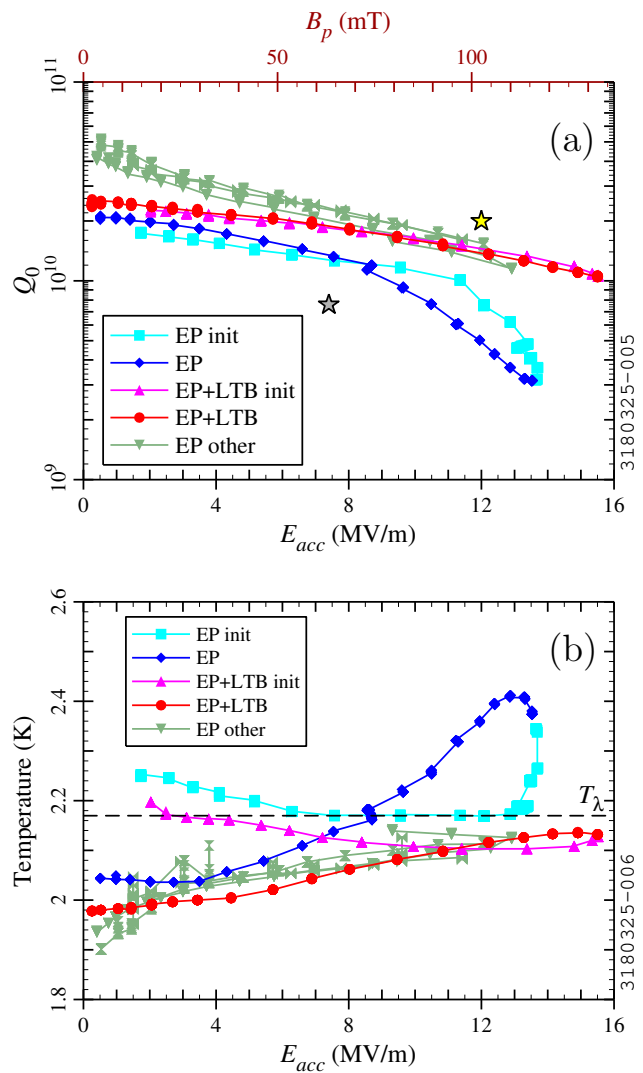

   \centering
   \GRAFlabelcoord{350}{255}
   \GRAFoffset{-65}{-58}
    \GRAFwidth[\figLHwidth]{490}{430}
   \incGRAFlabel{b530_epvbkix_e20_qebb}{(a)}\\
   \GRAFoffset{-65}{-65}
   \GRAFwidth[\figLHwidth]{490}{390}
    \incGRAFlabel{b530_epvbkix_e20_te}{(b)}\\
   \caption{(a) Measured quality factor and (b) corresponding temperature measurements as a function of accelerating gradient ($\Ea$) for $\beta = 0.53$ HWRs at $\sim 2$~K after EP and after additional LTB\@.  Cyan squares and magenta triangles: initial measurements on S53-159.  Blue and red: final measurements on S53-159.  Light green: additional cases.  Dashed line: $\lambda$ point temperature.  $B_p$ = peak surface magnetic field.}
   \label{fig:EPvLTB}
\end{figure}

We initially reported that EP produced HFQS for $\Ea  \geq 8$~MV/m and that it was mitigated by LTB, as seen in \cref{fig:EPvLTB}a (blue diamonds and red circles).  
However, as can be seen in \cref{fig:EPvLTB}, the bath temperature is not always as stable as desired during measurements at 2~K (for reasons discussed in \cref{S:CavTest}).  As seen in \cref{fig:EPvLTB}, initial measurements were done while the bath temperature was still decreasing (cyan squares and magenta triangles); at high cavity field, the bath temperature tended to increase.  In the test after EP (cyan squares and blue diamonds), the bath temperature stability was especially poor, possibly due to an issue with the cryogenic control loop.
This raises the question of whether the decrease in $Q_0$ for the EP'ed case is due to HFQS or merely the result of the increased bath temperature.
We evaluated the effect of the elevated temperature on $Q_0$ using the low-field measurements of $Q_0$ as a function of temperature, and found that the temperature alone could not account for the difference in $Q$-slope \cite{saito2023development}.  This assessment was based on the final round of measurements taken during the cold test (blue diamonds).  However, when we consider the initial measurements starting between 2.2 and 2.3~K (cyan squares) we observed significantly less HFQS for data points taken at lower temperatures.  Hence the HFQS after EP may be less severe than we initially reported in Ref.~\cite{saito2023development}.
 
 The difference between the cyan squares and blue diamonds cannot be accounted for based solely on the temperature dependence of $Q_0$ at low field, but may be the result of a jump in the $Q$-slope when the temperature increases beyond the $\lambda$ point temperature marking the transition to superfluid helium (dashed line in \cref{fig:EPvLTB}b).  An abrupt jump in the $Q$-slope at the $\lambda$ point has been reported in past tests on other cavity types, for example in Refs.~\cite{PRSTAB8:042003, NAPAC2011:TIP085}, and is consistent with measurements we have done on another FRIB HWR\@.  If we discount measurements taken above the $\lambda$ point, the results shown in \cref{fig:EPvLTB} do not provide very clear evidence of HFQS in the EP case.

Some additional tests were done after the initial studies reported in Ref.~\cite{saito2023development}, but without LTB\@.  The results are included in \cref{fig:EPvLTB} (light green).  The additional cases are (i) a repeat test on the same cavity (S53-159) after flux trapping studies, plastic deformation, and another round of light EP; (ii) a cavity which received half bulk BCP and half bulk EP (S53-155), with additional steps to reduce FE; (iii) a cavity which was mechanically polished after bulk EP (S53-124).  In all cases, an administrative limit was imposed at $\Ea \sim 12$~MV/m, so HFQS above that field is not ruled out.  However, the new results were obtained with better bath temperature stability and support the hypothesis that the apparent severe HFQS seen in the first measurements (blue diamonds) is the result of elevated bath temperature.  All 3 of the new cases show similar performance, in spite of the different cavity histories.  At low field, the new results show higher $Q_0$ than seen in the first round results, though this is within the spread in performance seen in past production cavities.  We note that the measured field emission X-rays were below 1~mR/hr for all of the measurements shown in \cref{fig:EPvLTB}.
Thus, despite several tests, the extent of HFQS for FRIB HWRs after EP alone is not fully quantified by our results so far.
 
\begin{figure}
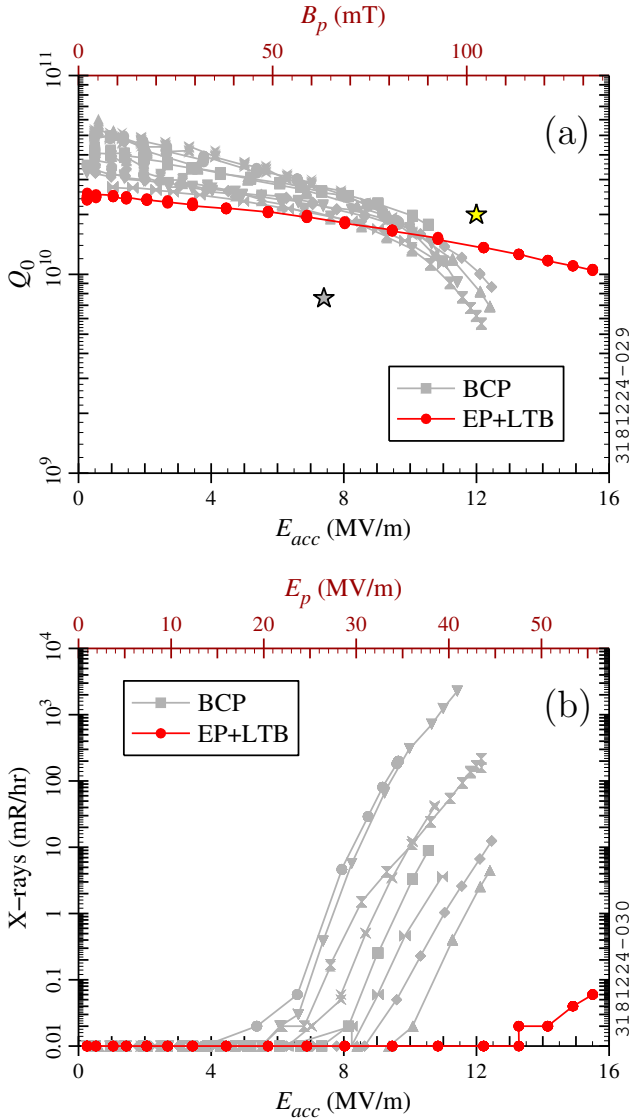

   \centering
   \GRAFlabelcoord{350}{255}
   \GRAFoffset{-65}{-58}
    \GRAFwidth[\figLHwidth]{490}{430}
    \incGRAFlabel{b530_517v_e20qebb}{(a)}\\
    \incGRAFlabel{b530_517v_e20xebb}{(b)}\\

    \caption{Comparison of measurements at $\sim 2$~K on $\beta = 0.53$ HWRs after BCP or EP+LTB: (a) quality factor and (b) X-rays as a function of gradient.  BCP cases: cavities certified and installed into a cryomodule (SCM517) near the end of FRIB production.  Silver star: goal for FRIB production cryomodules.  Gold star: goal for present effort.\label{fig:BCPvEPnLTBtyp}}
\end{figure}

 Further findings from our study include~\cite{saito2023development}
\begin{enumerate}
\item Post-EP LTB reduced the BCS surface resistance ($\Rbcs$) by a factor of 2; $\Rbcs$ was inferred by fitting the measured low-field surface resistance as a function of temperature~\cite{saito2023development}.  As discussed above, the reduction in $\Rbcs$ does not have a major impact on $Q_0$ at 2~K for 322 MHz.

\item EP and LTB are effective to produce high $Q_0$ at a high gradient of up to 15 MV/m without HFQS\@. The field was limited by quench. It should be emphasized that this is the first demonstration of the benefit of EP+LTB for low-frequency medium-$\beta$ cavities, similar to ILC cavities.

\item The quality factor at high gradient is greatly improved by EP+LTB, in comparison with cavities prepared with BCP as shown in \cref{fig:BCPvEPnLTBtyp}a.  However, as previously mentioned, $Q_0$ remains below our goal of $Q_0 \geq 2\times10^{10}$ at $\Ea = 12$~MV/m. 

\item Field emission X-rays were relatively low and lower than the average for FRIB production cavities prepared with BCP, as seen in \cref{fig:BCPvEPnLTBtyp}b.  We consider the smoother surface finish from EP to be an improvement over BCP for FE mitigation, though our results so far are not adequate to provide a statistical argument in favor of EP.
\end{enumerate}

\begin{figure}
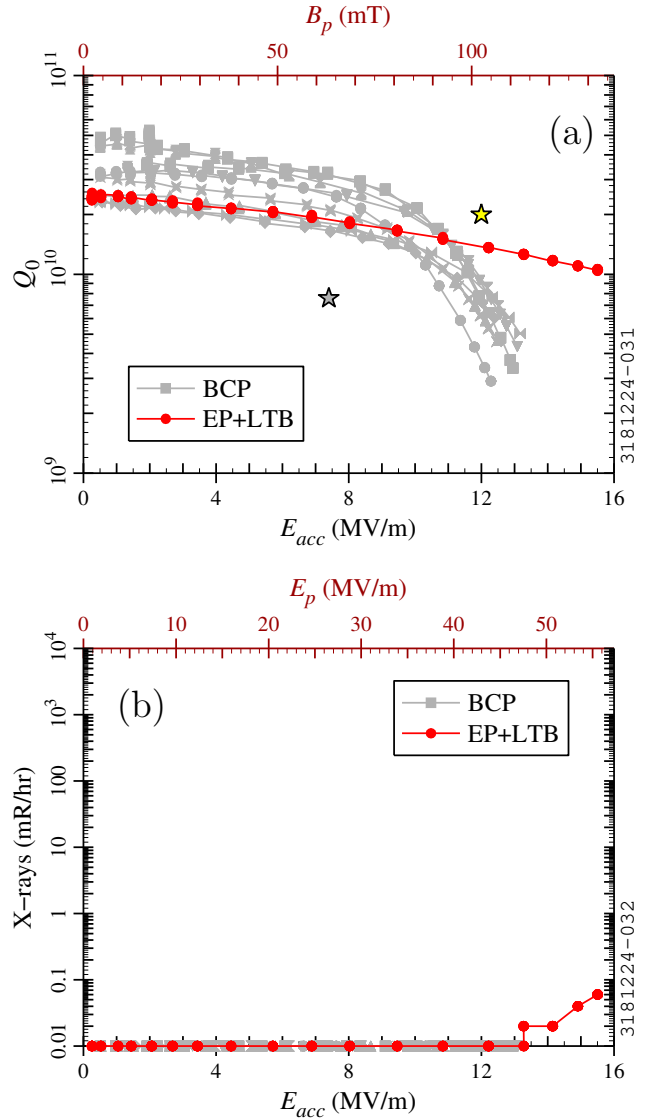

    \centering
    \GRAFlabelcoord{350}{255}
    \GRAFoffset{-65}{-58}
    \GRAFwidth[\figLHwidth]{490}{430}
    \incGRAFlabel{b530_lofev_e20qebb}{(a)}\\
    \GRAFlabelcoord{25}{255}
    \incGRAFlabel{b530_lofev_e20xebb}{(b)}\\

    \caption{Comparison of measurements at $\sim 2$~K on $\beta = 0.53$ HWRs after BCP or EP+LTB: (a) quality factor and (b) X-rays as a function of gradient.  BCP cases: FE-free cavities certified in the latter third of FRIB $\beta = 0.53$ production.}
    \label{fig:BCPvEPnLTBnoFE}
\end{figure}

The BCP'ed cavities in \cref{fig:BCPvEPnLTBtyp} (gray) were used for a FRIB cryomodule produced near the end of the production run.  The peak surface magnetic field ($B_p$) and peak surface electric field ($E_p$) are included in the upper axes in \cref{fig:BCPvEPnLTBtyp} for reference (we note that, typically, for a given gradient $\Ea$, lower-$\beta$ cavities have higher peak surface fields than higher-$\beta$ cavities).  All of the BCP'ed cavities showed FE X-rays at high gradient.  However, the decrease in $Q_0$ at high field is not due to field emission loading, as can be seen from \cref{fig:BCPvEPnLTBnoFE}, which compares the EP+LTB results to BCP'ed cavities which had no FE\@.  This shows that the BCP'ed cavities still have ``pure'' HFQS when they do not have field emission.  Hence the mitigation of HFQS via EP+LTB is not merely a side effect of FE reduction.

Comparing \cref{fig:EPvLTB} with \cref{fig:BCPvEPnLTBtyp} and \cref{fig:BCPvEPnLTBnoFE}, it is clear that the BCP'ed cavities have HFQS starting consistently at $\Ea \sim 10$~MV/m.  On the other hand, if we discount the results at elevated bath temperature, the EP'ed cavities do not show HFQS at $\Ea \sim 12$~MV/m, and the EP+LTB case shows no HFQS at $\Ea = 15$~MV/m, in clear contrast to the BCP cases.

\section{Experimental preparation: Flux Trapping Mitigation}
\subsection{Residual magnetic field\label{S:resid}}

\begin{figure}[tbp]
   \centering
   \includegraphics*[width=0.8\columnwidth]{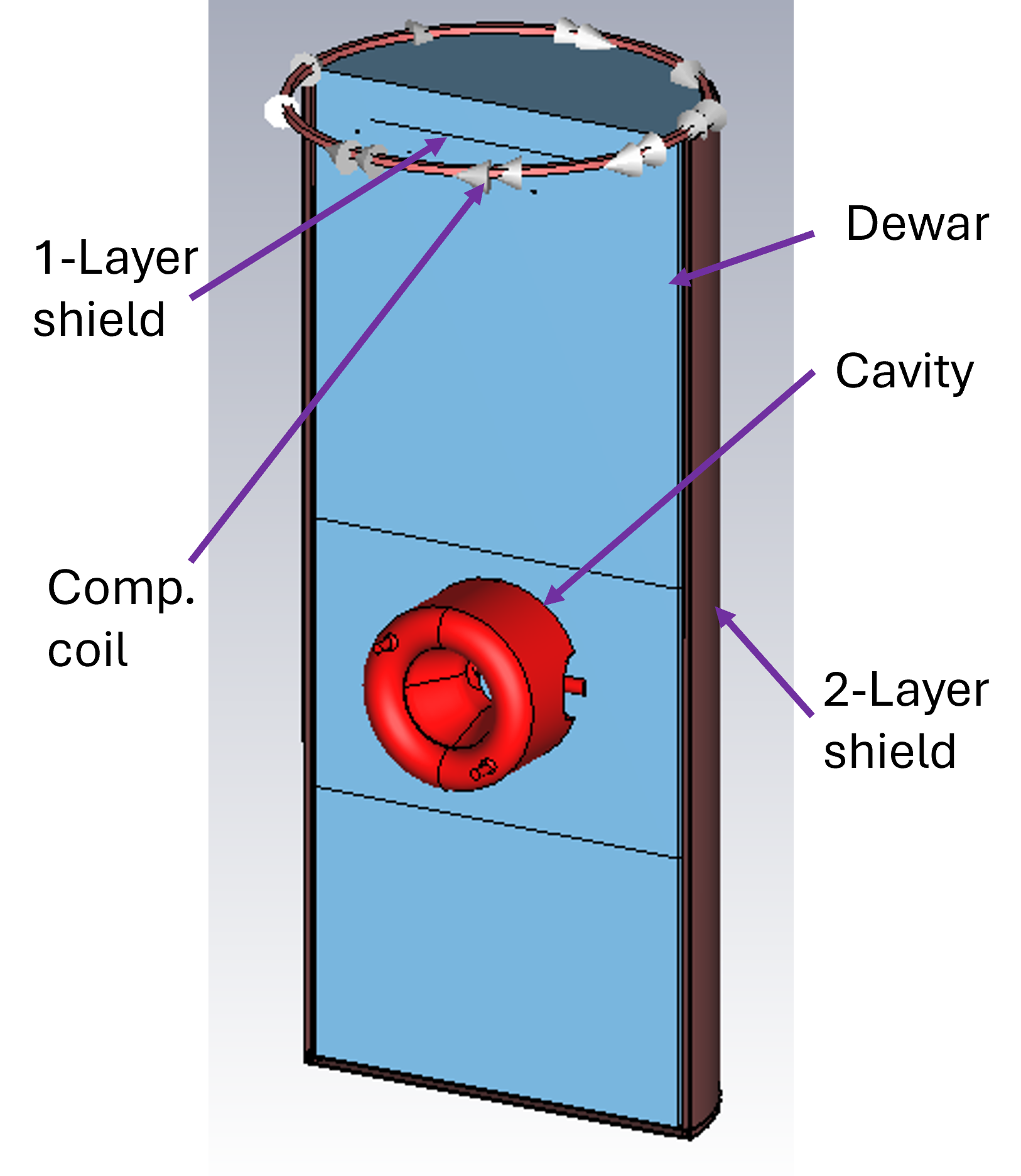}
   \caption{Drawing of a $\beta = 0.53$ HWR (red), test Dewar (blue), surrounding magnetic shield layers (brown), and  compensation coil (orange).  (The HWR's helium jacket is not shown.)}
   \label{fig:Dewar}
\end{figure}

\begin{figure}[tbp]
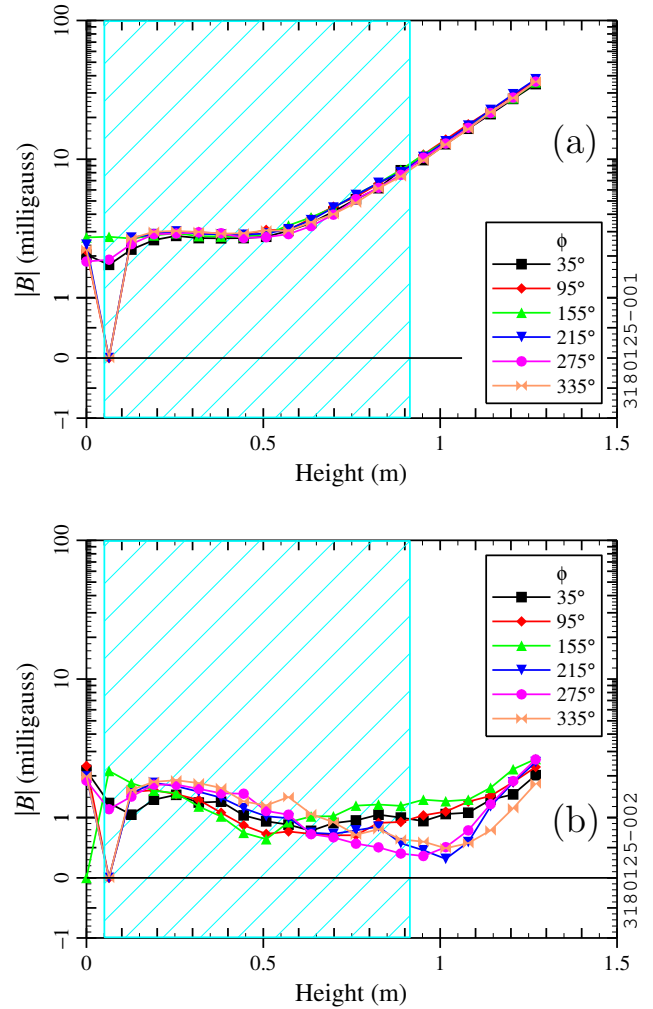

   \centering
   \GRAFoffset{-65}{-65}
   \GRAFwidth[\figLHwidth]{490}{390}
   \GRAFlabelcoord{350}{200}
   \incGRAFlabel{b_map_pit2_0}{(a)}\\
   \GRAFlabelcoord{350}{80}
   \incGRAFlabel{b_map_pit2_13da}{(b)}
   
   \caption{Measured magnetostatic field inside one of the FRIB Dewars (a) with the compensation coil off and (b) with the compensation coil at 1.3 A\@.  The field was measured at 6 different azimuthal angles ($\phi$) along a horizontal circle of diameter 0.4~m.  The light blue cross-hatching provides an approximate indication of the useful zone for cavity placement.}
   \label{fig:ResidFld}
\end{figure}

The FRIB test Dewars have a magnetic shield made of $\mu$-metal surrounding each Dewar. There are 2 layers of $\mu$-metal on the bottom and side walls (outside the Dewar) plus 1 layer of $\mu$-metal on the insert (inside the Dewar, below the lid; with some holes for vacuum piping, cryogens, cables, etc).  As a result, there is a gap in the shielding along the top edge of the Dewar. A compensation coil allows for partial cancellation of the leakage field (\cref{fig:Dewar}).  \Cref{fig:ResidFld}a shows a recent measurement of the magnetic field in a Dewar at room temperature with the compensation coil off.   For these measurements, the HWR was not present and 3 magnetometers were used to measure each of the field components.  The goal for certification testing of FRIB production cavities (tested without the compensation coil) was a residual field $< 15$~milligauss. As seen in \cref{fig:ResidFld}a, the measured field is 2 to 8~mG in the useful zone for cavity placement (light blue cross-hatching); we estimate that the field is 2 to 5~mG at the HWR location.  The compensation coil allows for an overall decrease in the residual field (\cref{fig:ResidFld}b); we will return to this point in \cref{S:afc}.  Additional information from earlier measurements of the residual fields in the FRIB Dewars can be found in previous papers \cite{saito2023development, PRAB24:112003, NIMA1059:168985}.

\subsection{Local magnetic shield design\label{S:lmgs}}

\begin{figure}
   \centering
   \includegraphics*[width=\columnwidth]{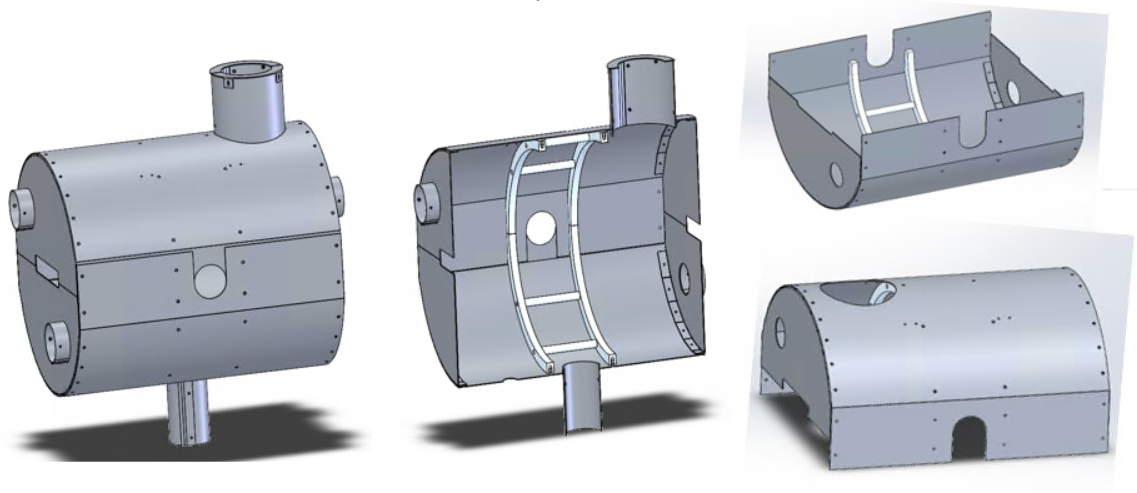}
   \caption{Right: upper and lower portions of the LMGS\@.  Middle: sectional view of the LMGS with sleeves.  Left: LMGS assembly.}
   \label{fig:LMGS}
\end{figure}

A local magnetic shield made of permalloy was designed to reduce the residual field to $\sim 0.1$~milligauss, in order to effectively eliminate the power dissipation due to flux trapping. As shown  in \cref{fig:LMGS} and \cref{fig:LMGSpic}, the LMGS surrounds the helium jacket. The design and analysis of the shield was done using the CST Studio Suite \cite{ICAP2006:THM2IS03}.  The upper port in the jacket (helium outlet) had the most field leakage, which necessitated the addition of a $\mu$-metal sleeve to reach the desired residual field. Similar sleeves were added for the RF input port and rinse ports. With the sleeves, the field decreased from $\sim 5$~mG outside to $\sim 0.1$~mG at the cavity, as shown in \cref{fig:LMGSfldCST}. Measurements on the as-fabricated shield at room temperature showed an attenuation factor of 99.6\% on the center axis, which should correspond to a residual field of 0.02~mG in the Dewar.


\begin{figure}[tbp]
    \centering
    \GRAFwidth[\figHLLwidth]{708}{746}
    \GRAFlabelcoord{150}{50}
    \GRAFlabelbox{70}{60}
    \incGRAFboxlabel{F6a}{(a)}
\ifthenelse{\lengthtest{\columnwidth=\textwidth}}
{\relax}%
{\\[1.5ex]}
    \GRAFwidth[\figFwidth]{220}{304}
    \GRAFlabelcoord{0}{0}
    \GRAFlabelbox{25}{20}
    \incGRAFboxlabel{F6b}{(b)}
    \caption{(a) LMGS isometric view showing the ports and sleeves. (b) LMGS assembled onto a $\beta = 0.53$ HWR (S53-159) installed on a test insert.}
    \label{fig:LMGSpic}
\end{figure}   

\begin{figure}[tbp]
    \centering
    \GRAFwidth[\figSwidth]{1402}{715}
    \GRAFlabelcoord{1200}{450}
    \GRAFlabelbox{140}{120}
    \incGRAFboxlabel{F7a}{(a)}\\[2ex]
    \GRAFwidth[\figSwidth]{534}{367}
    \GRAFlabelcoord{464}{0}
    \GRAFlabelbox{70}{45}
    \incGRAFboxlabel{F7b}{(b)}
    \caption{Field attenuation of the LMGS from CST simulation results: (a) field along a vertical path through the upper sleeve and the sleeveless bottom opening; (b) 2D section showing the field leakage from 2 ports with sleeves and 3 sleeveless openings.}
    \label{fig:LMGSfldCST}
\end{figure}   

\subsection{Mitigation of flux trapping from thermo-currents}

\begin{figure}[tbp]
   \centering
   \includegraphics*[width=\columnwidth]{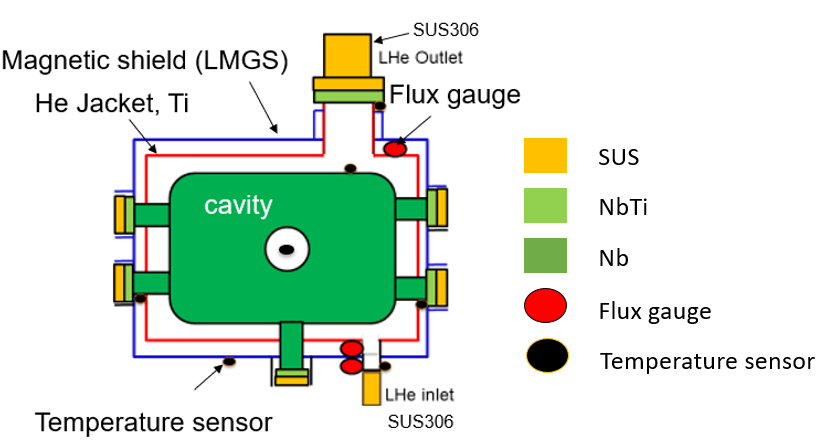}
   \caption{Schematic of the HWR with LMGS, temperature sensors (black ellipses), and flux gate magnetometers (red ellipses).}
   \label{fig:schema}
\end{figure}

As illustrated in \cref{fig:schema}, FRIB cavities are in the presence of several joints between different materials during a cold test: the cavity is made from Nb, the helium jacket is made from Ti, the cavity flanges are NbTi, and the liquid helium inlet and outlet lines are stainless steel (SUS). When a large temperature difference ($\geq$10 K) occurs during cool-down, the Seebeck effect will produce thermoelectric currents if there is a path for a current loop. A thermo-current will produce a magnetic field, which will in turn produce flux trapping if present during the cavity's transition from normal conducting to superconducting (at $T_c$ = 9.25~K). This effect can be eliminated by strategically-placed insulators to prevent current loops, but this cannot easily be done for FRIB HWRs. Hence, a simpler approach, using uniform cooling (UC) to reduce the temperature difference, was investigated instead.

As shown \cref{fig:schema}, a number of sensors were installed to quantify the effect of thermo-currents: 7 temperature sensors; and 3 flux gate magnetometers (flux gauges), 2 on the helium jacket and 1 on the outside of the LMGS\@. The magnetic sensors measure one field component only; they were oriented to measure the vertical and one horizontal component.

\subsubsection{Fast cool-down}

Fast cool-down (FC) was typically used for FRIB production cavity tests. It allows cavities to be cooled down within 45 to 90 minutes without careful attention to temperature gradients. Using the sensors described above, the measured temperature difference between 2 sensors on bottom ports is greater than 10~K, as seen in \cref{fig:CoolTvt}a (blue and red curves).

\subsubsection{Uniform cooling}

\begin{figure}[bp]
    \centering
    \includegraphics[width=\figXLwidth]{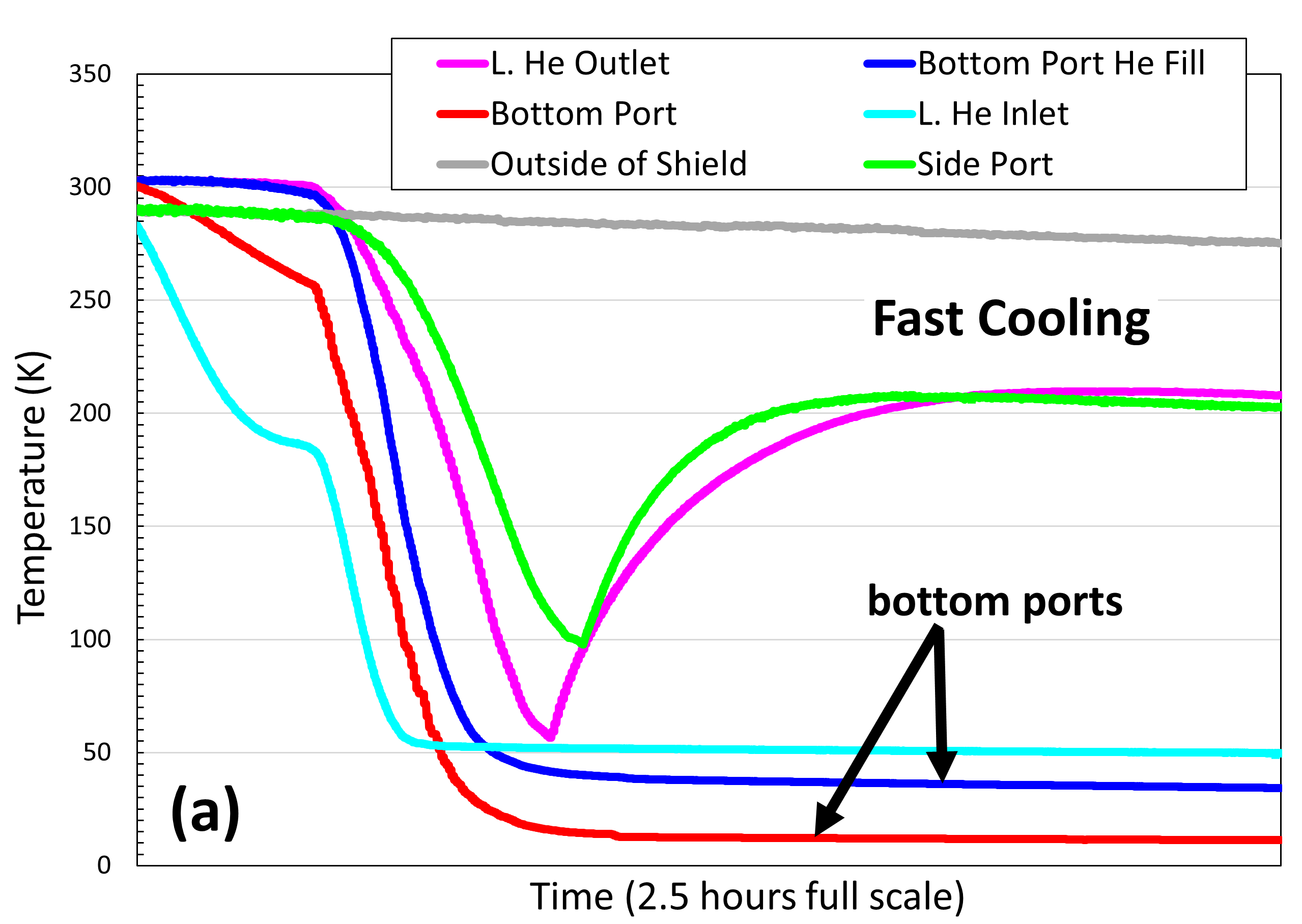}
    \includegraphics[width=\figXLwidth]{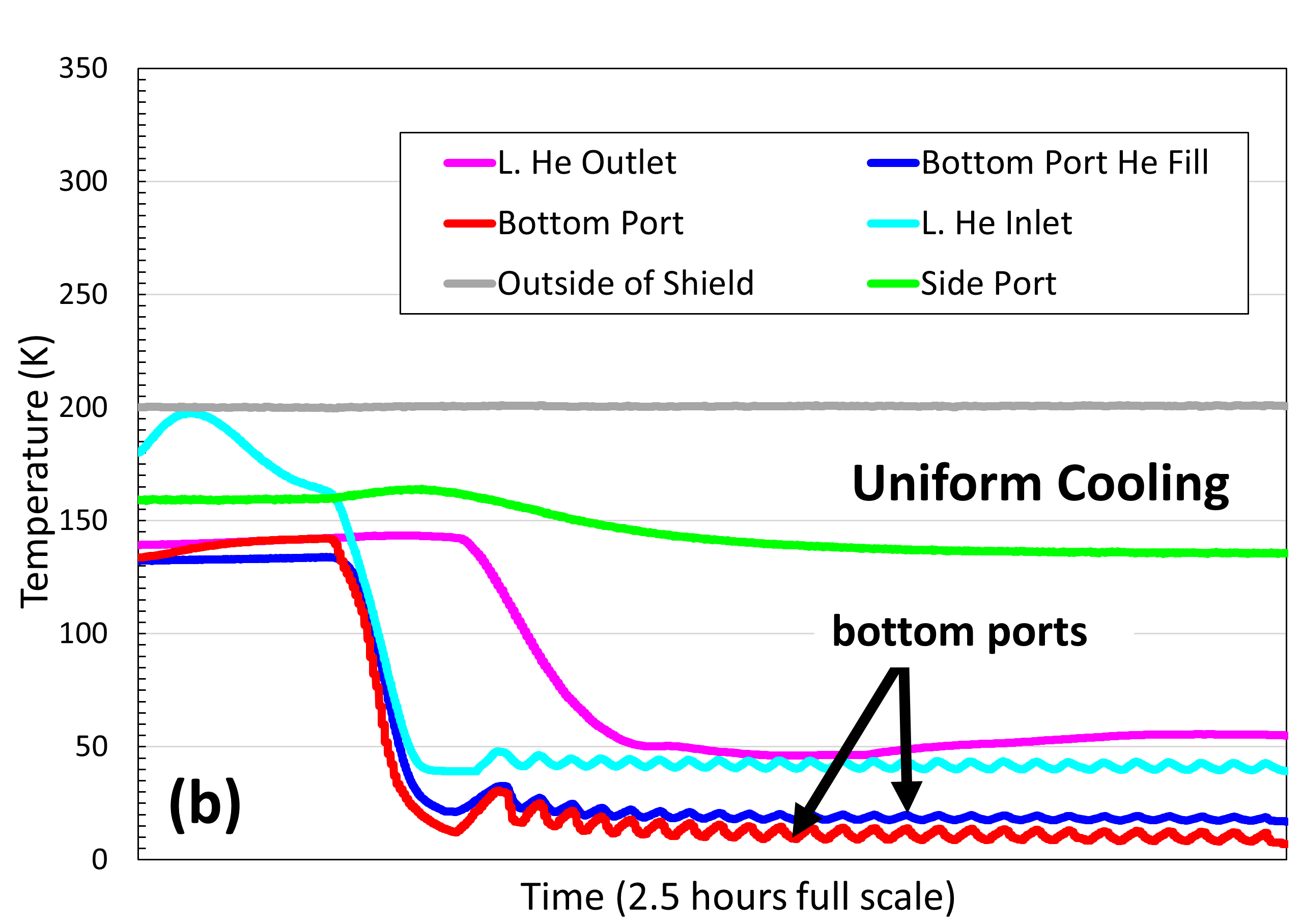} 
    
    \caption{Measured temperatures as a function of time during cool-down of an HWR: (a) fast cool-down; (b) uniform cooling.}
    \label{fig:CoolTvt}
\end{figure}

In contrast to fast cool-down, in which there are large temperature gradients, uniform cooling (UC) aims to minimize temperature differences. For our cold test system, this can be done by repeatedly opening and closing the liquid helium supply valve when the cavity temperature is near $T_c$. The cavity can be cooled rapidly to near $T_c$ before cycling the supply valve. An example of uniform cooling is shown in \cref{fig:CoolTvt}b. In this case, the blue and red curves show that the temperature difference between the bottom ports is within 10~K.

\subsection{Active field cancellation\label{S:afc}}

\begin{figure}[bp]
    \centering
    \includegraphics[width=\figXLwidth]{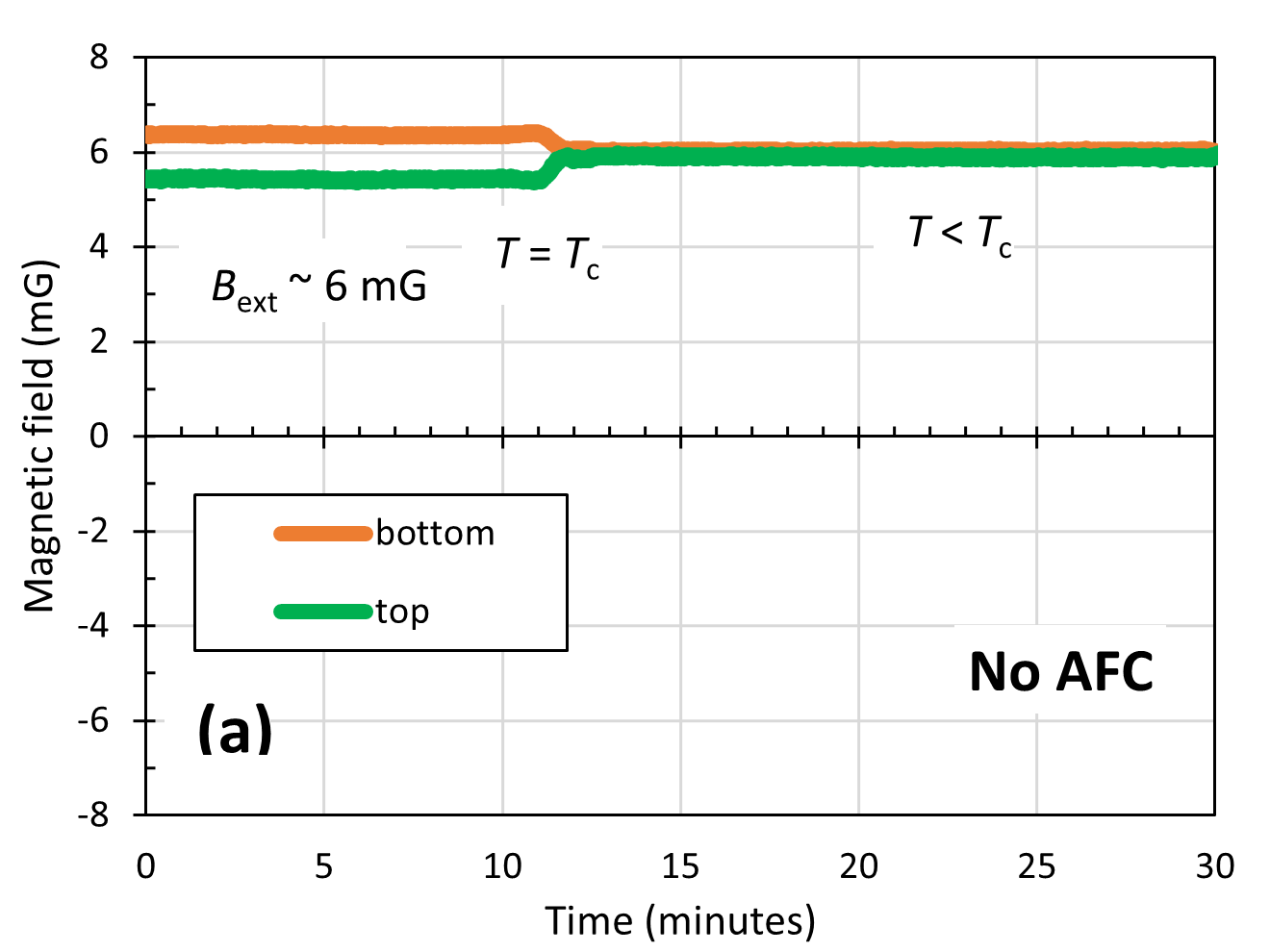} 
    \includegraphics[width=\figXLwidth]{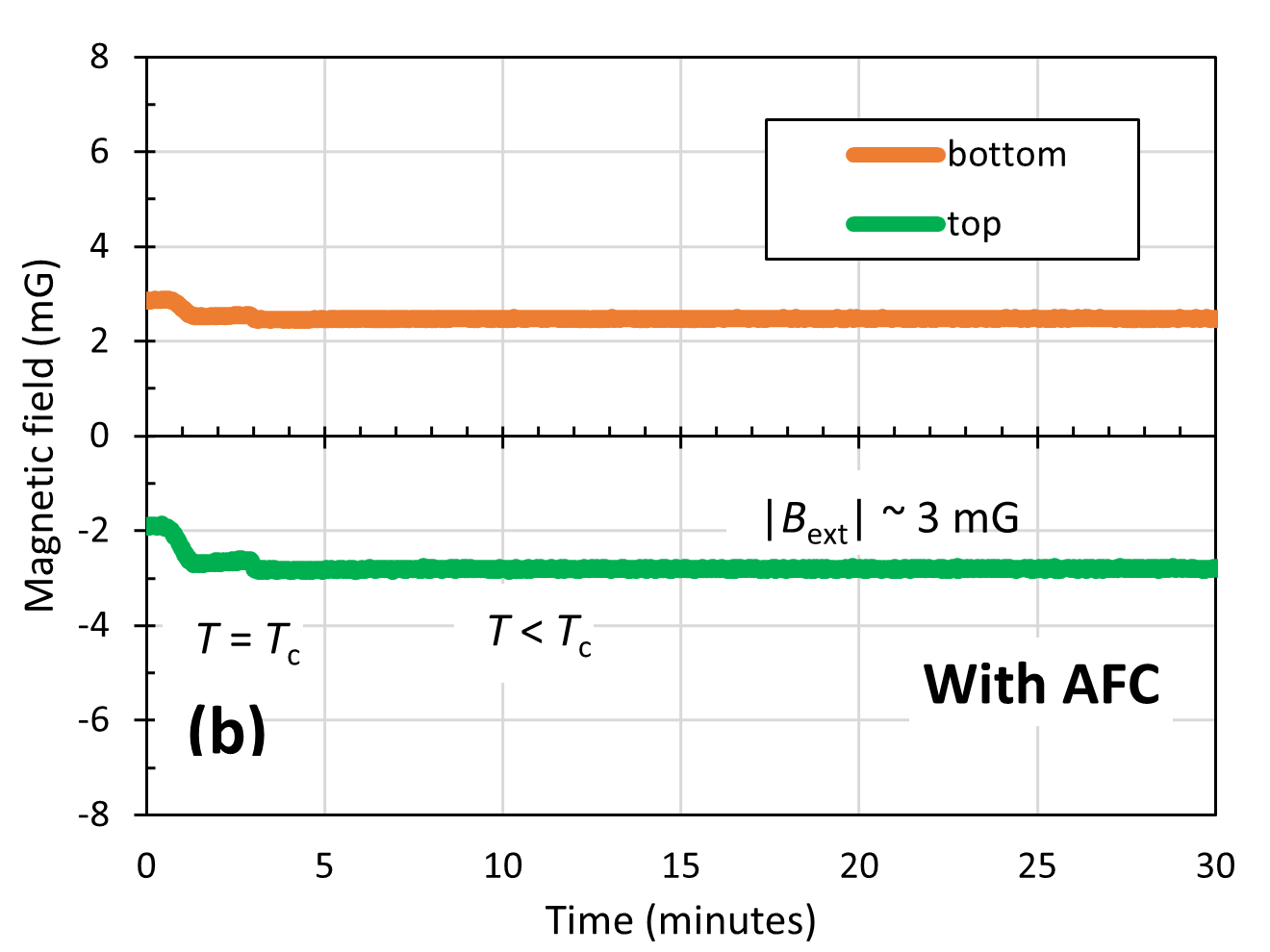}
    
    \caption{Measured magnetic field as a function of time with uniform cooling: (a) without AFC and (b) with AFC.}
    \label{fig:AFC}
\end{figure}

Active field cancellation (AFC) can reduce the residual magnetic field without additional magnetic shielding. AFC is done by applying a dc current to a compensation coil at the top of the Dewar (\cref{fig:Dewar}). The magnetic field from the coil counteracts the leakage of the ambient field into the Dewar from the gap in the shield around the top of the Dewar. The coil we used was originally developed for studies on prototype cavities for the proposed FRIB energy upgrade \cite{NIMA1059:168985}.  A recent measurement of the magnetic field with a coil current of 1.3 A is shown in \cref{fig:ResidFld}b.  From \cref{fig:ResidFld}, we see that the compensation coil can reduce the residual field in the relevant zone from a magnitude of 2 to 5~mG to $\leq 2$~mG at room temperature.
Combining AFC and UC, we can reduce the residual field from 6~mG to $\pm 3$~mG as the cavity transitions through $T_c$, as shown in \cref{fig:AFC}.
Thus, the residual field is still higher in the presence of cryogens and thermo-currents than measured at room temperature.

\section{Results: Flux Trapping Mitigation\label{S:res}}

To study the effect of flux trapping, we did Dewar tests for 4 different combinations: (A) FC with LMGS, (B) UC with LMGS, (C) UC without LMGS, and (D) UC without LMGS but with AFC.  

\subsection{Local magnetic shield with fast cool-down}

\begin{figure}[tbp]
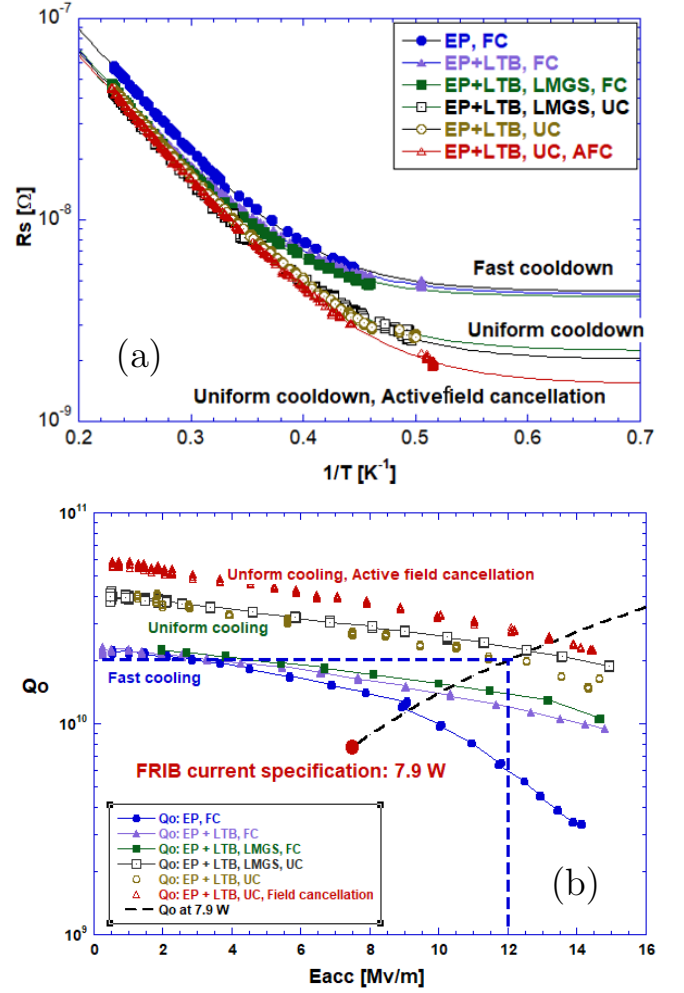

   \centering
    \GRAFwidth[\figLHwidth]{600}{460}
    \GRAFlabelcoord{75}{100}
    \GRAFlabelbox{70}{50}
    \incGRAFboxlabel{F11}{(a)}
    \GRAFlabelcoord{475}{75}
    \GRAFlabelbox{75}{50}
    \incGRAFboxlabel{F12}{(b)}
   \caption{Measurements for all HWR cold tests for Objectives 1 and 2: (a) RF surface resistance at low field ($\Ea \approx 2$~MV/m) as a function of inverse temperature; (b) measured quality factor at $\sim 2$~K as a function of gradient.}
   \label{fig:RsQo}
\end{figure}

We measured the RF surface resistance $R_s$ as a function of temperature at $\Ea \approx 2$~MV/m, as shown in \cref{fig:RsQo}a, without and with the LMGS installed.  With FC, the measured $R_s$ was very similar without (purple triangles) and with the LMGS (green squares). The fitted value of residual resistance ($\Rres$) decreased by a small amount, from 4.4 n$\Omega$ without the LMGS to 4.3 n$\Omega$ with the LMGS. 

Measurements of the quality factor as a function of gradient at $\sim 2$~K are shown in \cref{fig:RsQo}b.
For the FC cases at high gradients, $Q_0$ is slightly lower without the LMGS (purple triangles) and slightly higher with the LMGS (green squares).
We note that the LMGS does not mitigate flux trapping from thermo-currents.

\subsection{Local magnetic shield with uniform cooling}

Measurements of the combination of UC and LMGS are included in \cref{fig:RsQo} (hollow black squares).
The $\Rres$ decreases to 2.2 n$\Omega$ with this combination.  The quality factor is significantly higher at both low and high fields,
reaching $2.3\times10^{10}$ at $\Ea = 12$~MV/m, thus slightly exceeding our project goal.

\subsection{Uniform cooling without local magnetic shield}

For a more clear picture, additional measurements were done without the LMGS\@.
For this combination, the measured residual field was 6~mG near $T_c$, as seen in \cref{fig:AFC}a. The results are again included in \cref{fig:RsQo} (hollow black circles); there is very little difference in $R_s$ resulting from the removal of the LMGS: the black circles overlap with the back squares. Without the LMGS, the fitted $\Rres$ is 2.0 n$\Omega$, a bit smaller than the combination of UC + LMGS, but this difference is probably not statistically significant. At high field, $Q_0$ is slightly smaller without the LMGS, but this may not be statistically significant.

\subsection{Uniform cooling with active field cancellation}

With UC, no LMGS, and active field cancellation, the magnitude of the measured residual field is 3~mG near $T_c$ as seen in \cref{fig:AFC}b above. Hence we expect less flux trapping and lower losses. As seen in \cref{fig:RsQo} (hollow red triangles), $\Rres$ decreases to the lowest of our measured values, 1.5 n$\Omega$ and $Q_0$ reaches our highest measured value, $2.8\times10^{10}$ at $\Ea = 12$~MV/m.  The latter value exceeds our project goal by a comfortable margin.

\section{Discussion}
\subsection{Residual resistance: dependence on residual field}

\begin{figure}[hbp]
   \centering
   \includegraphics*[width=\columnwidth]{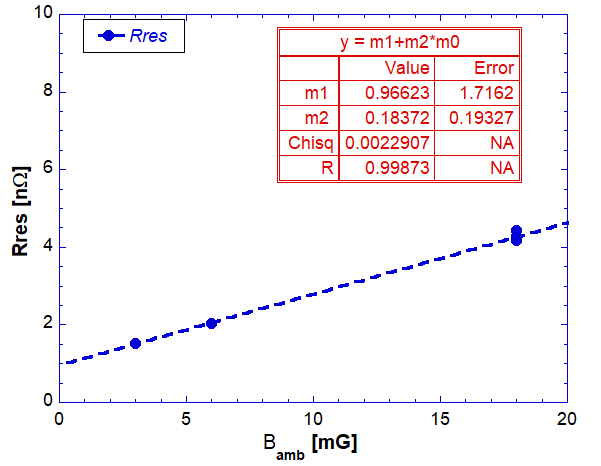}
   \caption{Fitted values of $\Rres$ for different measurements on the HWR as a function of the residual magnetic field measured near $T_c$ and linear fit.}
   \label{fig:RresvB}
\end{figure}

We measured the residual magnetic field near $T_c$ for each combination.  We measured 18~mG for FC without LMGS, 6~mG for UC without LMGS, and 3~mG for UC and AFC. As seen in \cref{fig:RresvB}, our results indicate a linear relationship between the residual surface resistance $\Rres$ and the residual magnetic field $\Bext$.  Hence we can express the field-dependent term by a linear equation: 
\begin{equation}
R_\mathrm{mg} = C \sqrt{f} \Bext \, = 0.184 \, \Bext\, .\label{eq:Rmg}
\end{equation}
We include the RF frequency $f$ in the coefficient based on the frequency dependence of the surface resistance for a normal-conducting metal.  Although, as discussed in \cref{S:TrapMitig}, the sensitivity to trapped flux varies depending on cavity shape, material properties, and other considerations, the coefficient of 0.184 (for $\Bext$ in milligauss and $R_\mathrm{mg}$ in n$\Omega$) is reasonably consistent with the value of 0.35 reported in the literature for EP'ed and BCP'ed cavities at 1500~MHz~\cite[Section 5.1]{hsp2023:srfta:saet} and the value of 0.433 for a 1300~MHz cavity measured at KEK~\cite{SRF1999:TUP005}.

The fitted line in \cref{fig:RresvB} suggests that $\Rres$ would be 1 n$\Omega$ with zero residual magnetic field; the corresponding value of $Q_0$ would be
 $7.2\times10^{10}$ for $\Ea = 2$~MV/m at 2~K.  

\subsection{\texorpdfstring{$Q$}{Q}-slope\label{S:Qslope}}

As seen in \cref{fig:RsQo}b, a clear slope in $Q_0$ as a function of field is observed in all of our measurements, even if for a field-emission-free cavity. As discussed in \cref{S:HiFldIntro}, thermal feedback (TF) is a possible cause for MFQS, though other models have been proposed.  The TF model considers the temperature difference between the helium bath and the inner wall of the cavity associated with conduction heat transfer of the RF dissipated power through the cavity wall with finite thermal conductivity and heat transfer across the Nb-He interface.  Assuming a linearized relationship between $R_s$ and the temperature $T$ a temperature-independent thermal conductivity, and a constant interface resistance, the thermal feedback model can be used to solve for the dependence of $Q_0$ on $\Ea$.  The result can be expressed as
\begin{equation}
\frac{1}{Q_0} = c_1 + c_2 \Ea^2\, ,\label{eq:ThFBQE}
\end{equation}
with $c_0$ and $c_1$ being constant in the linearized case.  For completeness, a basic derivation of \cref{eq:ThFBQE} is provided in the appendix.

\begin{figure}[bp]
   \centering
   \includegraphics*[width=\columnwidth]{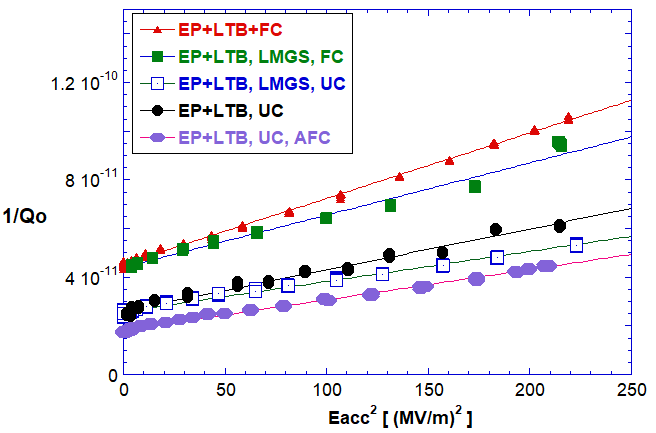}
   \caption{Measurements of quality factor at $\sim 2$~K as a function of gradient for HWRs tests: inverse of $Q_0$ as a function of gradient squared.}
   \label{fig:Qinv}
\end{figure}

To assess the usefulness of the TF model, we plot $1/Q_0$ as a function of $\Ea^2$ in \cref{fig:Qinv}. All of the measurements can be fit reasonable well by a linear relationship between $1/Q_0$ and $\Ea^2$. (The highest-field data point in the FC + LMGS case is perhaps an exception.) Hence the linearized TF model may reasonably explain our $Q$-slope measurements.  Although TF model fits the data well, we cannot rule out other contributions to the $Q$-slope based on these measurements.

In order to lessen the $Q$-slope, $c_2$ must decrease.  According to the TF model, this requires a smaller $\Delta T$ between the inner wall of the cavity and the helium bath and hence an increase in the thermal conductivity and/or interface resistance. If the TF model is indeed correct model, an increase in the thermal conductivity will mitigate the $Q$-slope and allow HWRs to reach higher $Q_0$ values at higher fields.

\subsection{Field emission}

The field emission onset was relatively high in all of the tests of the EP'ed HWR, with no X-rays seen up to $\Ea \approx 12$~MV/m. The measured X-rays were less than 0.1~mR/hr below 15~MV/m (\cref{fig:BCPvEPnLTBtyp}), which provides a comfortable margin relative to our project goals. However, additional FE onset margin would be needed for higher gradient operation (such as $\Ea >  15$ MV/m).  As was seen in \cref{fig:BCPvEPnLTBtyp}, the FE onset of our EP'ed HWR is significantly higher than that of typical BCP'ed HWRs. We can argue that the smoother surface produced by EP is responsible for the higher FE onset, but we cannot provide statistical evidence for this based on EP of one HWR\@. We plan to acquire more experience with EP on HWRs while producing spare and refurbished HWR cryomodules for the FRIB linac.  Other possibilities for further improvement in FE performance may come from future cavity assembly procedures using robotics~\cite{la1900cleanroom} or optimization of cavity high-pressure water rinsing.   

\subsection{Ultimate FRIB HWR performance}

\begin{figure}[tbp]
   \centering
   \includegraphics*[width=\columnwidth]{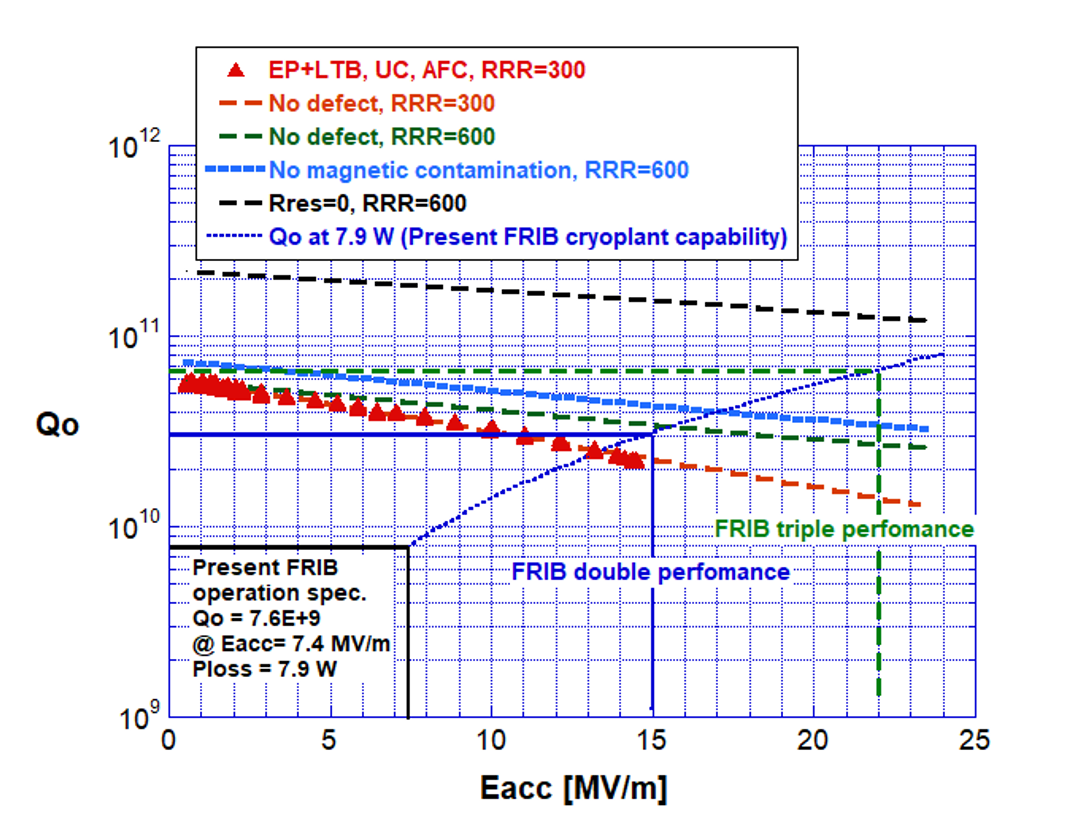}
   \caption{Quality factor as a function of gradient for the FRIB $\beta$ = 0.53 HWR: measured values at 2 K and theoretical curves to illustrate the ultimate performance potential.}
   \label{fig:ultQ}
\end{figure}

If we assume a superheating field $B_\mathrm{sh}$ of 200~mT for niobium~\cite{valles2011superheating}, the maximum gradient for the FRIB $\beta$ = 0.53 HWR would be 23~MV/m based on the current RF design parameters for which $B_p/\Ea$ = 8.6~mT/(MV/m). \Cref{fig:ultQ} considers the potential cavity performance under different assumptions. The present capacity of the FRIB cryoplant is 7.9 W/cavity (dotted blue curve in \cref{fig:ultQ}). If the $Q$-slope is dominated by thermal feedback and if the thermal conductivity of the cavity wall were doubled, operation at 15~MV/m would be theoretically possible (\cref{fig:ultQ}, dashed green curve). If, in addition, the residual magnetic field could be completely eliminated, operation at 17~MV/m would be possible in principle (\cref{fig:ultQ}, dashed light blue curve). In this scenario, increasing the thermal conductivity of the Nb wall and complete elimination of the residual field provide the path for high gradient operation with $\Ea > 15$~MV/m; with these assumptions and measures, operating the $\beta$ = 0.53 HWRs at double the present gradients is theoretically possible even with the present cryoplant capacity.

More speculatively, tripling of the gradient would be challenging but may not be completely excluded. To achieve a goal this ambitious, the residual surface resistance of 1 n$\Omega$ (which we extrapolate as being still present with zero residual field) would have to be further reduced (in \cref{fig:ultQ}, the dashed black curve corresponds to the limit $\Rres \rightarrow 0$). A further reduction in $\Rres$ would likely require a better understanding of the residual loss mechanisms other than trapped magnetic flux.  A residual resistance as low as 0.5~n$\Omega$ has been reported at 1.3~GHz with very high Nb purity \cite{SRF2001:MA008}, which suggests that achieving $\Rres < 1$~n$\Omega$ at 322~MHz may not be impossible, though doing so may indeed be a challenge.  Similar $\Rres$ values have been reported in more recent studies of medium-temperature baking \cite{PRAPPL13:015024}.

Nitrogen doping/infusion or MTB may be another method to achieve higher quality factors for HWRs.  As described in \cref{S:Dopeintro}, the experience so far with N-doping has been mainly with high-frequency elliptical cavities, with limited results at frequencies for lower frequencies; however, the available information suggests that the 2~K performance of 322~MHz HWR could be improved with N-doping or N-infusion.  Additionally, as mentioned in \cref{S:TransProc}, we are investigating a wet nitrogen doping technique \cite{LINAC2024:TUAA002}, which differs from the N-doping methods developed for ILC-style cavities, with additional results to be published in a separate paper.

We must add that increasing the operational gradient may pose challenges beyond the fundamental limits of cavity performance. Useful operation at higher field would entail higher power transfer from the cavity to the beam, and might require additional RF drive power.  Control of the cavity amplitude and phase would be more challenging as well,
since Lorentz force detuning increases quadratically with the accelerating gradient.

\section{Summary}

Electropolishing and low-temperature baking were applied to a FRIB $\beta = 0.53$ half-wave resonator in lieu of buffered chemical polishing used for FRIB cavity production. With EP and LTB, we reached an accelerating gradient of 15 MV/m, which is about twice the design goal for FRIB production cryomodules.
The field was limited by thermal breakdown (``quench''). The thermal breakdown suggests that surface imperfections may still be present, such that techniques such as centrifugal barrel polishing prior to EP may be beneficial.

At 1.3~GHz, low-temperature baking is needed to remove high-field $Q$-slope.  Our results suggest that HFQS may not appear after EP in 322 MHz resonators, but further experimental confirmation is needed for a definitive conclusion.  However, our results show conclusively that EP and LTB are effective to improve the high-gradient performance of medium-$\beta$ 322~MHz cavities, consistent with well-established results for high-$\beta$ 1.3 GHz cavities.

The quality factor is improved by uniform cooling and active field cancellation to mitigate thermoelectric currents and decrease the residual magnetic field.  With UC and AFC, we reached $Q_0$ = $2.8\times10^{10}$ at $\Ea = 12$ MV/m, which is almost 3 times the $Q_0$ goal at 160\% of the $\Ea$ goal for FRIB production cryomodules.  We find that flux trapping by the thermo-currents accounts for 80\% of the residual surface resistance ($\sim 5$~n$\Omega$) seen in $\beta = 0.53$ Dewar tests with fast cool-down. Thus, to reach a high $Q_0$, mitigation of thermo-currents is essential. This is an important finding for the design of future high-gradient cryomodules. Numerous joints with different materials are typically used in cryomodules, so the thermo-current mitigation may require significant design adjustments. The addition of insulating layers may be a way to prevent current loops, but would be challenging from a technical point of view.

Dynamic load measurements on FRIB HWR cryomodules during bunker tests and in the FRIB tunnel indicated that the cryogenic losses were lower than would be expected from the $Q_0$ values measured in cavity Dewar tests~\cite{xu2021completion}. Our investigation of thermo-currents and cooling methods suggest an explanation for this observation: likely there was more flux trapping in Dewar tests of production HWRs due to fast cool-down and there is less flux trapping in the cryomodules because they are cooled down more slowly.  Further mitigation of thermo-currents may be needed in order to operate at higher gradients.

In spite of the improvement in the high-field $Q_0$, a slope in $Q$ as a function of field was still observed in all of the tests with EP and LTB\@.  The imperfect thermal conductivity of the cavity wall is a possible contributor to the $Q$-slope.  In this scenario, if the thermal conductivity were doubled, we could in principle reach $Q_0 = 4\times10^{10}$ at $\Ea = 15$~MV/m. Hence operation at $\Ea = 15$~MV/m is possible in principle, if compatible with the thermal breakdown limit.
For operation at even higher gradients ($\Ea > 15$~MV/m) with the present FRIB cryoplant capacity, we anticipate that better mechanical surface preparation, improved thermal conductivity of the cavity wall, and full mitigation of thermoelectric currents would be essential.

\section*{Acknowledgments}

Our work has been a collaborative effort with the FRIB
cryogenics team, the FRIB cavity preparation team, and
the rest of the FRIB laboratory.
We especially appreciate the support work by Joseph Asciutto (inspection and mechanical polishing), David Norton (cavity test preparation and support for magnetic field measurements), John Schwartz (frequency and coupling measurements), Brian Barker (cavity preparation), Pete Donald (clean assembly), Sean Moskaitis (magnetic field measurements), and Patrick Tutt (magnetic field modeling).  We thank Jie Wei and Thomas Glasmacher for their ongoing support for this project.

Operation of the Facility for Rare Isotope Beams is supported by the U.S. Department of Energy Office of Science, Office of Nuclear Physics under Award Number DE-SC0023633. This work was supported by the U.S. Department of Energy Office of Science, DE-S RC114424.  The authors are grateful for this financial support.

\appendix

\section*{Appendix: Linearized thermal feedback model}

Using the definition of the quality factor (see, for example, Ref.~\cite[Section 2.2.1]{padamsee1998rfsca}), we can write
\begin{equation}
\frac{1}{Q_0} = \frac{\Ploss}{\omega U}\, ,\label{eq:1/Qo}
\end{equation}
where $\Ploss$ is the dissipated power on the SRF cavity surface, $\omega$ is the cavity's resonant angular frequency ($2\pi f$), and $U$ is the stored energy, which is proportional to $\Ea^2$.  The surface resistance $R_s$ depends on the temperature $T_S$ of the cavity inner wall, which in general is not equal to the helium bath temperature $T_B$.  To linear order in $T$, we can write
\begin{equation}
R_s(T_S) = R_s(T_B) + \frac{\partial R_s}{\partial T} \Big|_{T=T_B} \Delta T\, ,\label{eq:Rs}
\end{equation}
where $\Delta T = T_S - T_B$.  The power dissipation
$\Ploss$ depends on $R_s$ and the surface RF magnetic surface field $H_s$ via an integral over the cavity inner surface:
\begin{eqnarray}
\Ploss &=& \frac{1}{2} R_s(T) \, \int H_s^2 \, dS \nonumber
\\ 
&=& \left[ \frac{1}{2} R_s(T_B) + \frac{1}{2} \frac{\partial R_s}{\partial T} \Big|_{T=T_B} \Delta T \right] \, \int H_s^2 \, dS\ \nonumber
\\
&=& (a + b \Delta T) \int H_s^2 \, dS\, .\label{eq:PlossRs}
\end{eqnarray}
The equations above are approximate, as in general the temperature $T$ may have some variation over the cavity surface, in which case $R_s(T)$ should, strictly speaking, remain inside the integral.  
A rigorous approach would be to allow the temperature and surface resistance to have a nonuniform distribution over the cavity surface, with
the quality factor representing the average RF surface resistance weighted according
to the surface magnetic field distribution.
A method to relate the local/intrinsic surface resistance to the weighted average surface resistance for an SRF cavity has been developed,
though homogeneous surface properties are assumed \cite{PRAB21:122001}.  This could in principle be generalized to the case of a nonhomogeneous surface resistance determined by the surface temperature distribution.  However, for simplicity,
we consider the temperature to be averaged over the cavity surface as an approximation.

Conduction heat transfer allows the RF power deposited on the inner cavity surface to travel to the liquid helium bath.  In steady state, we can write
\begin{equation}
\Ploss = \kappa S \, \frac{\Delta T}{t} = \kappa S \, \frac{(T_B - T_S)}{t}\, ,       
\end{equation}
where $S$ is the effective surface area, $t$ is the thickness of the cavity wall, and $\kappa$ is the thermal conductivity.
Solving for $\Delta T$, we obtain
\begin{equation}
\Delta T = \frac{t} {\kappa S} \, \Ploss \label{eq:DT}
\end{equation}
and therefore, since $\Ploss$ is proportional to $\Ea^2$, $\Delta T$ is proportional to $\Ea^2$.  We can combine Eq.~(\ref{eq:PlossRs}) and \cref{eq:DT} to obtain
\begin{equation}
\Ploss = c'_1 \Ea^2 + c'_2 \Ea^4\, ,\label{eq:Ploss}
\end{equation}
where the coefficients $c'_1$ and $c'_2$ are independent of $\Ea$.  However, in order for $c'_2$ to be truly constant, the thermal conductivity $\kappa$ must be a constant and hence we must neglect its dependence on the temperature.

Using \cref{eq:1/Qo}, and making use of the fact that $U$ is proportional to $\Ea^2$,  we can obtain an expression for $1/Q_0$ as a function of $\Ea$,
\begin{equation}
\frac{1}{Q_0} = \frac{\Ploss}{\omega U} = c_1 + c_2 \Ea^2\, ,\label{eq:ThFBQEa}
\end{equation}
consistent with \cref{eq:ThFBQE} in \cref{S:Qslope}.  As we have seen, we must linearize the dependence of the surface resistance on temperature and must assume a temperature-independent thermal conductivity in order for the coefficients $c'_2$ and $c_2$ to be constants independent of $\Ea$.

A more general derivation of the thermal feedback model would take into account convection heat transfer from the outer wall of the cavity to the helium bath or Kaptiza conductance at the interface \cite[Section 3.3.2]{hsp2009:rfsc:sta}, but with suitable assumptions, the end result is still consistent with \cref{eq:ThFBQEa}.

\bibliography{ref}

\end{document}